\documentclass[conference]{IEEEtran}
\IEEEoverridecommandlockouts
\usepackage[letterpaper, left=0.75in, right=0.75in, top=0.75in, bottom=1in]{geometry}
\usepackage{cite}
\usepackage{amsmath,amssymb,amsfonts}
\usepackage{textcomp}
\usepackage{xcolor}

\usepackage{graphicx}
\usepackage{stmaryrd}
\usepackage[ruled,vlined]{algorithm2e}
\usepackage{array}
\usepackage{stfloats}
\usepackage{url}
\usepackage{verbatim}
\usepackage{stmaryrd}
\usepackage{multirow}
\usepackage{booktabs}
\usepackage{makecell}
\usepackage{threeparttable}
\usepackage{xhfill}  

\usepackage{amsthm}
\newtheorem{theorem}{Theorem}
\newtheorem{definition}{Definition}

\usepackage[acronym]{glossaries}
\makeglossaries
\newacronym{idx}{SLS-\textsc{index}}{\underline{s}ecure \underline{l}earned \underline{s}patial index}
\newacronym{RQ}{SLRQ}{\underline{SL}S-\textsc{index}-based \underline{R}ange \underline{Q}ueries}
\newacronym{sbp}{SBP}{\underline{s}ecure \underline{b}ucket \underline{p}rediction}
\newacronym{spe}{SPE}{\underline{s}ecure \underline{p}oint \underline{e}xtraction}

\usepackage{bm}
\usepackage{subfigure}
\def\BibTeX{{\rm B\kern-.05em{\sc i\kern-.025em b}\kern-.08em
    T\kern-.1667em\lower.7ex\hbox{E}\kern-.125emX}}
\begin{document}

\addtolength{\textheight}{-0.07in}

\title{Towards Privacy-Preserving Range Queries with Secure Learned Spatial Index over Encrypted Data
}


\author{\IEEEauthorblockN{Zuan Wang\IEEEauthorrefmark{2},
		Juntao Lu\IEEEauthorrefmark{2}, 
		Jiazhuang Wu\IEEEauthorrefmark{2},
		Youliang Tian\IEEEauthorrefmark{3}\IEEEauthorrefmark{1},
		Wei Song\IEEEauthorrefmark{2},
		Qiuxian Li\IEEEauthorrefmark{4}, and
		Duo Zhang\IEEEauthorrefmark{5}}
	\IEEEauthorblockA{\IEEEauthorrefmark{2}School of Artificial Intelligence and Computer Science, Jiangnan University, Wuxi, China}
	\IEEEauthorblockA{\IEEEauthorrefmark{3}College of Big Data and Information Engineering, Guizhou University, Guiyang, China}
	\IEEEauthorblockA{\IEEEauthorrefmark{4}College of Big Data Engineering, Kaili University, Kaili, China}
	\IEEEauthorblockA{\IEEEauthorrefmark{5}School of Mathematics and Statistics, Guizhou University, Guiyang, China}
	\IEEEauthorblockA{Email: zuanwang@jiangnan.edu.cn, \{6243111044, 6243111053\}@stu.jiangnan.edu.cn,\\ yltian@gzu.edu.cn, songwei@jiangnan.edu.cn, \{qiuxianLL, sxzd0816\}@163.com}
	\thanks{$^{*}$Youliang Tian is the corresponding author.}}

\maketitle

\begin{abstract}
With the growing reliance on cloud services for large-scale data management, preserving the security and privacy of outsourced datasets has become increasingly critical. While encrypting data and queries can prevent direct content exposure, recent research reveals that adversaries can still infer sensitive information via access pattern and search path analysis. However, existing solutions that offer strong access pattern privacy often incur substantial performance overhead. 
In this paper, we propose a novel privacy-preserving range query scheme over encrypted datasets, offering strong security guarantees while maintaining high efficiency. To achieve this, we develop \gls{idx}, a secure learned index that integrates the Paillier cryptosystem with a hierarchical prediction architecture and noise-injected buckets, enabling data-aware query acceleration in the encrypted domain. 
To further obfuscate query execution paths, \gls{RQ} employs a permutation-based secure bucket prediction protocol. Additionally, we introduce a secure point extraction protocol that generates candidate results to reduce the overhead of secure computation.
We provide formal security analysis under realistic leakage functions and implement a prototype to evaluate its practical performance. Extensive experiments on both real-world and synthetic datasets demonstrate that \gls{RQ} significantly outperforms existing solutions in query efficiency while ensuring dataset, query, result, and access pattern privacy.
\end{abstract}


\section{Introduction}
With the rapid proliferation of cloud computing services, an increasing number of enterprises and organizations are outsourcing sensitive datasets to cloud platforms to leverage their scalable storage and powerful computational capabilities. However, the inherent trust issues and vulnerabilities of cloud infrastructures have raised significant concerns regarding data security and privacy. For instance, the 2023 T-Mobile data breach, which compromised the personal information of 37 million users, was caused by attackers exploiting vulnerabilities in third-party cloud APIs to access private data, such as addresses and dates of birth. Such incidents highlight the risks associated with storing unencrypted data on cloud platforms. Furthermore, attackers may further infer highly sensitive information, such as medical records or location trajectories, through sophisticated access pattern analysis.

\begin{figure}[t]
	\centering
	\includegraphics[width=8cm]{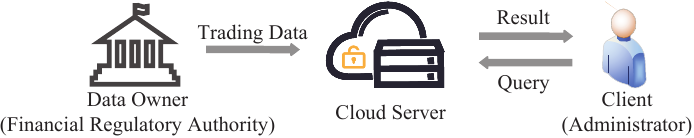}\vspace{-2mm}
	\caption{Example of range queries on the cloud platform.} 
	\label{exp_c}
	\vspace{-5mm}
\end{figure}

As illustrated in Fig. \ref{exp_c}, a financial regulatory authority outsources sensitive securities trading data, which comprises transaction timestamps, trader identifiers, asset symbols, trade volumes, and prices, to a cloud platform in order to support real-time analytical services for market surveillance and regulatory compliance. The public cloud enables SQL-style queries such as:
\texttt{select * from trades where price $\geq$ 100 and volume $\leq$ 5000}
to retrieve trade records matching specific conditions.
To ensure the confidentiality of both the data and queries, a straightforward solution is to encrypt the dataset and query predicates before outsourcing them to the cloud \cite{du2020privacy}, \cite{zheng2022secskyline}, \cite{wang2022efficient}. However, as demonstrated in \cite{islam2012access}, cloud providers may still infer sensitive information by analyzing the access patterns of encrypted queries, particularly when prior domain knowledge is available. Therefore, privacy protection should extend beyond basic encryption to include the following essential guarantees \cite{du2020privacy}, \cite{zheng2022secskyline}, \cite{lei2020fast}:
(1) Data privacy, to protect the content of the outsourced dataset;
(2) Query privacy, to conceal the query predicates;
(3) Result privacy, to ensure confidentiality of the returned results;
(4) Access pattern privacy, as emphasized in recent studies \cite{wang2022efficient, cui2020svknn}, to prevent the cloud from discerning which records satisfy a given query.

Motivated by this, our objective is to propose a secure range query protocol over encrypted datasets that achieves the aforementioned privacy guarantees while ensuring high efficiency. To protect data confidentiality, we employ the Paillier cryptosystem \cite{paillier1999public} to encrypt the datasets, queries, and results.
Unlike weaker primitives such as order-preserving encryption \cite{shen2021practical} or attribute-preserving encryption \cite{naveed2015inference}, Paillier ensures strong security guarantees and has been widely adopted in secure query processing \cite{du2020privacy, cui2020svknn, ding2021efficient}, with proven commercial applicability. Moreover, our index design is tailored to exploit Paillier’s homomorphism and can be seamlessly extended to support more advanced homomorphic encryption schemes.

However, secure protocols based on Paillier suffer from excessive query latency. To overcome this limitation, we propose a novel index structure that integrates a strong encryption (i.e., the Paillier cryptosystem) with data-aware optimization, thereby significantly reducing the volume of encrypted data that should be scanned during query processing.
Traditional index structures (e.g., R-tree-based schemes \cite{wang2013secure}) support efficient range queries but fail to leverage data distribution for performance gains. In contrast, learned index structures like Lisa \cite{li2020lisa} uses machine learning models to optimize data layout, but they operate only on plaintext data.
As a result, they are unsuitable for encrypted environments and risk leaking sensitive information such as datasets, queries, and results. 
To address this, we propose \gls{idx}, a secure learned index that maps encrypted data points into a rank space and uses space-filling curves \cite{qi2020effectively, sagan2012space} to preserve data locality and capture distributional patterns. This design enables privacy-preserving yet efficient query processing over encrypted datasets. Specifically, \gls{idx} adopts a hierarchical structure consisting of a head predictor, intermediate predictors, and leaf predictors, tightly integrated with secure multilayer perceptrons. The leaf predictor encrypts all model parameters using Paillier to ensure access pattern privacy, while the head and intermediate predictors encrypt only partial parameters, striking a balance between security and efficiency. Additionally, \gls{idx} supports secure index updates while preserving index privacy.

Despite these protections, cloud servers may infer dataset distributions, query contents, or result privacy by analyzing query access paths or result positions during the query processing, thereby compromising data confidentiality. For example, suppose an adversary knows some prior knowledge (e.g., known several data points and their corresponding positions on the dataset). If the adversary observes that the query covers these known data points, (s)he may infer the general content of this query by analyzing results and non-result points from prior information, as the access patterns of these points are exposed. 
Additionally, during query execution, the cloud server may capture and analyze the search paths on \gls{idx}, thereby revealing query unlinkability (i.e., search patterns). That is, the cloud may track query paths to obtain the positions of data points. To address these challenges, our \gls{RQ} incorporates the following three aspects:

1) To preserve the indistinguishability of the index structure, \gls{idx} integrates the Paillier cryptosystem with a dummy bucket injection mechanism. Moreover, \gls{idx}'s hierarchical prediction architecture with fuzzy label mechanisms enhances query efficiency by capturing data distribution while protecting data confidentiality.

2) We design a novel \gls{sbp} protocol that combines permutation and perturbation techniques with dummy bucket injection to obfuscate the prediction process during secure index traversal, thereby protecting search pattern privacy.

3) By leveraging the inherent inaccuracy of prediction ranges, specifically their inclusion of non-result points (i.e., false positives), we design the \gls{spe} protocol to securely obtain candidate results while preserving access pattern privacy.

In summary, our contributions are outlined as follows:

1) We develop \gls{idx}, a novel secure learned index structure that combines the Paillier cryptosystem, a hierarchical prediction architecture with fuzzy label mechanisms, and dummy bucket injection.
By learning inherent data distributions, \gls{idx} enables efficient data fitting in the encrypted domain and further supports dynamic updates through a secure index update mechanism.

2) We design a suite of secure protocols to support \gls{RQ}, which leverages obfuscation techniques to enforce the aforementioned privacy guarantees, with a particular focus on protecting search patterns and access patterns.

3) We conduct a comprehensive security analysis of \gls{RQ} and evaluate its performance through extensive experiments on both real-world and synthetic datasets. The results show that \gls{RQ} delivers significant improvements in query efficiency while ensuring strong privacy guarantees.

\section{Preparation and Problem Statement}\label{pre}
\subsection{Paillier cryptosystem}
We adopt the Paillier cryptosystem \cite{paillier1999public} as the underlying cryptographic primitive, owing to its additive homomorphic properties and semantic security. For brevity, we denote Paillier ciphertexts as $E(\cdot)$ or $\llbracket \cdot \rrbracket$, supporting the following homomorphic operations:
\begin{equation}
	D_{sk} \left( \left( \llbracket x_0 \rrbracket \cdot \llbracket x_1 \rrbracket \right) \bmod N^2 \right) = (x_0 + x_1) \bmod \mathbb{N},
\end{equation}
\begin{equation}
	D_{sk} \left( \llbracket x_0 \rrbracket^{x_1} \bmod N^2 \right) = (x_0 \cdot x_1) \bmod \mathbb{N},
\end{equation}
where $x_0, x_1 \in \mathbb{Z}_N$, and $\mathbb{N}$ is the product of two large prime numbers. Here, $D_{sk}(\cdot)$ denotes the decryption function under the secret key $sk$.

\subsection{System Model}
\begin{figure}[t]
	\centering
	\includegraphics[width=7cm]{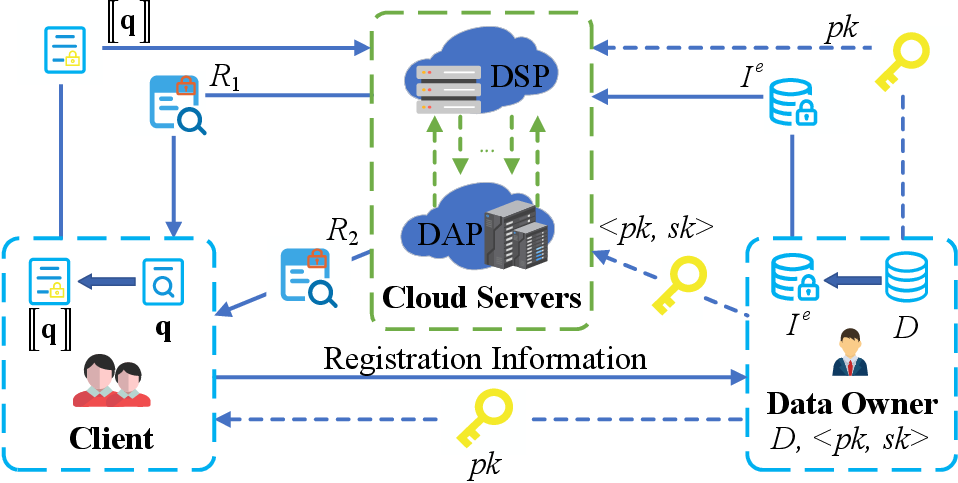}\vspace{-2mm}
	\caption{\label{tab1}System model.} 
	\label{sys}
	\vspace{-5mm}
\end{figure}

As illustrated in Fig.~\ref{sys}, our scheme employs two collusion-resistant cloud servers, a widely adopted architecture in secure query processing~\cite{du2020privacy,ding2021efficient,wang2022efficient}. 
These servers, referred to as the Data Service Provider (DSP) and the Data Assistance Provider (DAP), are typically regarded as independent and competitive entities, often operated by well-established companies such as Amazon or Microsoft, which makes collusion between them highly unlikely.
Alternatively, both servers can be part of the same reputable organization, with the DAP deployed as a high-security “vault” within the infrastructure. The roles of the entities involved are defined as follows:

1) \textbf{Data Owner (DO)}: As a trusted party, DO first generates a Paillier key pair \( \langle pk, sk \rangle \) via $\textbf{Setup}(1^{\lambda})$, distributing the public key \( pk \) to DSP and the full key pair  \( \langle pk, sk \rangle \) to DAP. The DO then constructs a secure index \( I^{e} \) using $\textbf{BuildIndex}(D, pk)$, and uploads \( I^e \) to the DSP for secure range query processing. To support updates, the DO employs the $\textbf{Update}$ algorithm to insert or delete the encrypted point $\llbracket{\bm{p}}\rrbracket$ from the index.

2) \textbf{Client}: As an authorized and trusted user (AU), the client generates an encrypted range query \( \llbracket {Q} \rrbracket \) using $\textbf{Trapdoor}(Q, pk)$, and sends  \( \llbracket {Q} \rrbracket \) to DSP. After query execution, the client receives partial results \( R_1 \) from DSP and \( R_2 \) from DAP, and combines them to derive the final result set \( R \).

3) \textbf{Cloud Servers}: Upon receiving the encrypted query \( \llbracket {Q} \rrbracket \), DSP, equipped with the public key \( pk \), employs the secure index \( I^e \) to compute encrypted results \( R^e \) via $\textbf{Query}(I^{e}, E(Q))$. To reduce dependency on the data owner during query execution, DAP, which holds the complete key pair \( \langle pk, sk \rangle \), assists in the secure computation. Once the query processing is completed, both the DSP and DAP use $\textbf{Return}(R^{e})$ to send their respective outputs to the client, who reconstructs the final results using obfuscation techniques. For further details on $\textbf{Return}(R^{e})$, please refer to our prior work \cite{wang2023efficient}.


\subsection{Problem Definitions}
\begin{definition}(\gls{RQ}).
	Given an encrypted dataset $\llbracket P \rrbracket = {\llbracket \bm{p}_0 \rrbracket, \ldots, \llbracket \bm{p}_{n-1} \rrbracket}$ and an encrypted range query $\llbracket Q \rrbracket = {\left[\llbracket q_{\text{l}}^{0} \rrbracket, \llbracket q_{\text{r}}^{0} \rrbracket\right], \ldots, \left[\llbracket q_{\text{l}}^{d-1} \rrbracket, \llbracket q_{\text{r}}^{d-1} \rrbracket\right]}$, where each encrypted point $\llbracket \bm{p}_i \rrbracket = {\llbracket \bm{p}_i^0 \rrbracket, \ldots, \llbracket \bm{p}_i^{d-1} \rrbracket}$ lies in a $d$-dimensional space, and $\langle\llbracket q_{\text{l}}^j \rrbracket$, $\llbracket q_{\text{r}}^j\rrbracket\rangle$ denote the encrypted lower and upper bounds of the query in the $j$-th dimension, respectively. Secure range queries with \gls{idx} (i.e., \gls{RQ}) retrieves all points within the specified query range $\llbracket Q \rrbracket$ while preserving the confidentiality of the dataset, query, results, and access patterns.
%
%
%
%
%
%
\end{definition}

\subsection{Security Model}\label{SecurityModel}
Both DSP and DAP are modeled as semi-honest entities~\cite{oded2009foun ,xu2021privacy,wu2020privacy}, meaning they follow the prescribed protocol but may attempt to infer sensitive information from encrypted data and intermediate results. We do not consider threats related to access control or insecure communication channels, as explored in~\cite{xu2018query}. Instead, we assume that the data owner (DO) is trusted, the user is authorized, and the communication channel is secure. Under this threat model, we define the security of our protocol using the well-established simulation-based paradigm~\cite{lindell2017simulate}, which has been widely adopted in secure computation~\cite{ding2021efficient,cui2020svknn,liu2015efficient}.

\begin{definition}(Security).  
	Let $\Pi = ($\textsf{Setup}, \textsf{BuildIndex}, \textsf{Trapdoor}, \textsf{Update}, \textsf{Query}, \textsf{Return}$)$ denote our \gls{RQ} protocol. Define leakage functions $\mathcal{L}_{\text{Build}}$, $\mathcal{L}_{\text{Update}}$, and $\mathcal{L}_{\text{Query}}$ that capture the information revealed during index construction, secure updates and query execution, respectively. Let $\mathcal{A}$ be a probabilistic polynomial-time (PPT) adversary and let $\mathcal{S}$ be a PPT simulator. Security is defined via the indistinguishability between the following two games:
	
	{\rm$\textsf{Real}_{\mathcal{A}}(\lambda)$:}  
	$\mathcal{A}$ selects a dataset $P$. The protocol runs $\textsf{Setup}(\lambda)$ to generate a Paillier key pair $\langle pk, sk\rangle$. Then, it computes $I^{e} \leftarrow \textsf{BuildIndex}(P, pk)/\textsf{Update}(I^{e}, {\bm p}_{new})$ and sends $I^{e}$ to $\mathcal{A}$. The adversary generates a polynomial number of queries $Q = \{Q_0, Q_1, \ldots, Q_{q-1}\}$. For each $Q_i$, the protocol runs $\textsf{Trapdoor}(Q_i, pk)$ to obtain the encrypted query $E(Q_i)$, which is used by $\mathcal{A}$ to execute $\textsf{Query}(I^{e}, E(Q_i))$ and obtain encrypted results $R^e$. Finally, $\mathcal{A}$ outputs a bit $b \in \{0, 1\}$ as the outcome of the game.
	
	{\rm$\textsf{Ideal}_{\mathcal{A}, \mathcal{S}}(\lambda)$:}  
	$\mathcal{A}$ choose a dataset $P^{*}$, and $\mathcal{S}$ generates a simulated index $I^*$ though $\mathcal{L}_{\text{Build}}$ and $\mathcal{L}_{\text{Update}}$, and then provides it to $\mathcal{A}$. The adversary issues queries $Q = \{Q_0, Q_1, \ldots, Q_{q-1}\}$. For each $Q_i$, $\mathcal{S}$ uses $\mathcal{L}_{\text{Query}}$ to simulate the encrypted query $E(Q_i^*)$ and sends it to $\mathcal{A}$. The adversary then obtains the simulated encrypted results $R^{e*}$ by executing $\textsf{Query}(I^*, E(Q_i^*))$, and finally outputs a bit $b \in \{0, 1\}$.
	
	$\Pi$ achieves $(\mathcal{L}_{\text{Build}}, \mathcal{L}_{\text{Update}}, \mathcal{L}_{\text{Query}})$-security if, for any PPT adversary $\mathcal{A}$, there exists a simulator $\mathcal{S}$ such that:
	$${\left| \Pr[\textsf{Real}_{\mathcal{A}}(\lambda) = 1] - \Pr[\textsf{Ideal}_{\mathcal{A}, \mathcal{S}}(\lambda) = 1] \right|} \leq \text{negl}(\lambda),$$
	where $\text{negl}(\lambda)$ denotes a negligible function in the security parameter $\lambda$.
\end{definition}

\section{Secure Learned Spatial Index}

\subsection{Core Secure Predictor}\label{csp}
To determine the order and storage location of data points, we adopt a Z-curve-based ordering strategy inspired by \cite{qi2020effectively}. Given \( n \) points \( P = \{ {\bm p}_1, {\bm p}_2, \dots, {\bm p}_n \} \) in a \( d \)-dimensional space, each point is mapped to a \textit{ranking space}, where its coordinates represent rank values across dimensions, with tie-breaking ensured by comparing subsequent dimensions. 
After that, each point is mapped to one-dimensional space via a Z-curve, with its curve value denoted as \( {\bm p}.\mathit{cur} \). Sorting by curve values yields a global order, from which points are grouped into fixed-size buckets of capacity \( b \). The bucket identifier is computed as \( {\bm p}.\mathit{bkt} = \lceil {\bm p}.\mathit{ord} / b \rceil \), where \( {\bm p}.\mathit{ord} \) denotes the sorted position. Each bucket stores encrypted points \( E({\bm p}) \) and the encrypted minimum bounding rectangle \( E(\bm{mbr}) \).

To securely predict the target bucket without revealing data contents, the bucket index for each result is computed as:
\begin{equation}
	E({\bm p}.bkt) = \mathcal{M}^{e}({\bm p}),
\end{equation}
where $\mathcal{M}^{e}$ denotes a model $\mathcal{M}$ with encrypted parameters, trained over the ranking space. To capture nonlinear patterns, we adopt a multi-layer perceptron (MLP) for $\mathcal{M}$ and minimize the $L_2$ loss using stochastic gradient descent (SGD):
\begin{equation}
	\label{loss}
	\mathcal{L}{oss} = \sum_{{\bm p}\in P}(\mathcal{M}({\bm p})-{\bm p}.bkt)^{2}.
\end{equation}
Training proceeds until the maximum prediction error,
\begin{equation}
	err_{max} = \max_{\forall {\bm p}\in P}\left\{\left| {\bm p}.bkt -\mathcal{M}({\bm p}) \right| \right\},
\end{equation}
falls below a predefined threshold.
Unlike typical training regimes, our indexing model does not aim to avoid overfitting. Instead, the objective is to achieve a perfect fit to the ranked values.



To facilitate secure computation and ensure data confidentiality, we adopt a three-layer MLP architecture with parameters encrypted via the Paillier cryptosystem. Based on this, we propose two variants: SMLP$_p$ with plaintext outputs and SMLP$_c$ with ciphertext outputs, offering a trade-off between efficiency and security within the indexing architecture. 
The inputs of both SMLP$_p$ and SMLP$_c$ are encrypted points $E({\bm p})$. In SMLP$_p$, only the hidden-layer bias $b^{(2)}$ is encrypted, while the weight matrix $W^{(2,1)}$ remains in plaintext with added noise to preserve security. The hidden output is computed homomorphically as $\psi = E({W^{(2,1)}}^T + b^{(2)})$, followed by decryption and ReLU activation. The output layer remains unencrypted to reduce computation. Note that SMLP$_p$ with noise still needs to be controlled to within $err_{max}$ of the maximum error. In contrast, SMLP$_c$ encrypts all parameters and computes $E({\bm p}.\mathit{bkt})$ entirely in the encrypted domain. Additionally, a secure ReLU ($\mathrm{SReLU}$) function is constructed using secure integer comparison (SIC) \cite{ding2021efficient}, defined as follows:
\begin{equation}
	\label{reluep}
	\mathrm{SReLU}(E(x)) = \begin{cases} 
		E(x),  & \mbox{if } \mathrm{SIC}(\llbracket 0 \rrbracket, \llbracket x \rrbracket)=1 \\
		E(0) & \mbox{if } \mathrm{SIC}(\llbracket 0 \rrbracket, \llbracket x \rrbracket) = 0
	\end{cases}.
\end{equation}


\begin{figure*}[t]
	\centering
	\includegraphics[width=15cm]{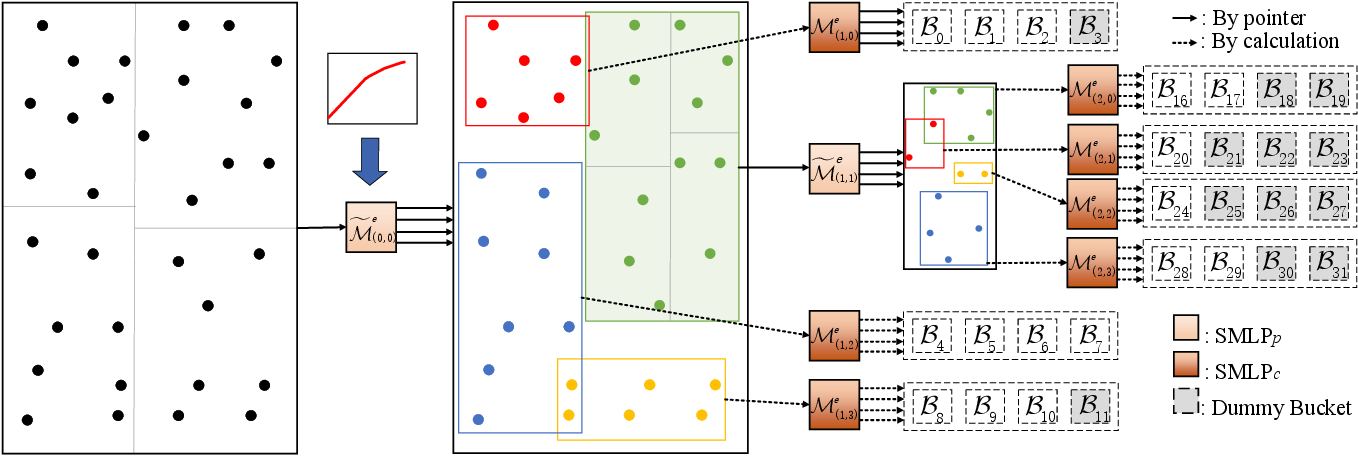}
	\vspace{-3mm}
	\caption{Structure of \gls{idx}. This secure index comprises a head predictor, intermediate predictors and leaf predictors. The head and intermediate predictors employ $\widetilde{\mathcal{M}}^{e}$, while leaf predictors use ${\mathcal{M}}^{e}$ to preserve data confidentiality. Moreover, to hide the privacy of search patterns, \gls{idx} also incorporates a fuzzy label mechanism along with carefully designed dummy predictors and buckets.}
	\label{myidx}
	\vspace{-3mm}
\end{figure*}

\subsection{Secure Index Architecture}\label{arc}
As shown in Fig. \ref{myidx}, we propose an indexing architecture based on secure MLPs, namely \gls{idx}, which integrates learned indexing with a hierarchical partitioning strategy. Given a dataset $P$ of size $n$, our \gls{idx} recursively partitions $P$ until each subset contains at most
$m$ encrypted points, which are suitable for accurate bucket prediction by $\mathcal{M}$. Depending on privacy needs, each partition is assigned either SMLP$_p$ or SMLP$_c$, which collectively form a secure index that is uploaded to the DSP. For ease of presentation, we describe \gls{idx} using $d=2$, which readily extends to any $d \in \mathbb{N}^{+}$.

The DO divides the dataset $P$ into approximately $\left \lceil m/b \right \rceil$ partitions, enabling reuse of the model $\mathcal{M}$ to predict partition IDs. 
This is achieved by constructing \( 2^{\lfloor \log_4(m/b) \rfloor} \times 2^{\lfloor \log_4(m/b) \rfloor} \) grids that aligns with the data distribution. Specifically, the data space is split into \( 2^{\lfloor \log_4(m/b) \rfloor} \) vertical columns (based on x-coordinates), each containing roughly \( \left\lceil \frac{n}{2^{\lfloor \log_4(m/b) \rfloor}} \right\rceil \) points. Each column is then further divided horizontally (by y-coordinates) into the same number of rows, resulting in grid cells with at most \( \left\lceil \frac{n}{4^{\lfloor \log_4(m/b) \rfloor}} \right\rceil \) points. A space-filling curve (SFC) of appropriate order is applied to assign a curve value to each cell. A model \( \mathcal{M}_{0,0} \) is then trained to map any point \( p \in P \) to the curve value of its corresponding cell. This model reuses the structure and loss function of \( \mathcal{M} \), but instead of predicting \( p.\mathit{bkt} \), it predicts the curve value derived from the grid. To preserve data confidentiality, the DO applies SMLP$_p$ to model $\mathcal{M}_{(0,0)}$, which is accordingly referred to as the head predictor and denoted by $\widetilde{\mathcal{M}}^{e}_{(0,0)}$.

Next, \( \mathcal{M}_{(0,0)} \) is used to predict curve values and form partitions $\{pt_{(0,1)},pt_{(0,2)},...,pt_{(0,k)}\}$, where $k$ denotes the maximum number of predictors allowed at each level. If a partition still contains more than the threshold $m$ points, the grid-based partitioning strategy is recursively applied, and further models \( \mathcal{M}_{(1,j)} \) are trained using the curve values derived from the grid. To preserve data confidentiality, the DO also applies SMLP$_p$ to intermediate models $\mathcal{M}_{(i,j)}$ ($i \neq 1$), which are accordingly referred to as intermediate predictors and denoted by $\widetilde{\mathcal{M}}^{e}_{(i,j)}$.

This recursive process continues until partitions contain fewer than $m$ points. At that stage, leaf models $\mathcal{M}_{(i,j)}$ are trained to predict the exact bucket ${\bm p}.bkt$ for each point $\bm p$ in the partition. To preserve data confidentiality, the DO applies SMLP$_c$ to these leaf models $\mathcal{M}_{(i,j)}$ ($i \neq 1$), which are thus referred to as leaf predictors and denoted by ${\mathcal{M}}^{e}_{(i,j)}$.

To mitigate search pattern leakage, we introduce a fuzzy label mechanism. Instead of returning a direct leaf predictor ID, the DSP uses a vector \( \mathcal{F} \) composed of encrypted indicators (i.e., \( \llbracket 1 \rrbracket \) and \( \llbracket 0 \rrbracket \)) to compute the correct leaf predictor:

\begin{equation}
	\mathcal{M}^{e}_{(h-1,j)} = \prod_{k=1}^{\eta} \mathrm{SM}(\mathcal{F}_{j}[k], \mathcal{M}^{e}_{(h-1,k)}),
\end{equation}
where \( \eta \) is the number of child predictors of \( \widetilde{\mathcal{M}}^{e} \), and $\mathrm{SM}$ denotes secure multiplication \cite{wang2023efficient}.

To defend against inference based on bucket counts, artificial noise buckets and points are introduced. Each leaf node is padded with encrypted dummy data (e.g., \( \llbracket 0 \rrbracket \)) to fill each bucket to capacity, while simultaneously ensuring both query correctness and efficiency are preserved.

\subsection{Update}
We also introduce update mechanisms for the \gls{idx} stored in encrypted cloud environments, where general update operations are reduced to fundamental insertions and deletions.

{\bf Insertion.} The Do maintains a plaintext \gls{idx} and the corresponding dataset $D$ locally. To insert a new point $\bm p$ into $D$ and update \gls{idx}, the DO first computes the target bucket position using the prediction model $\mathcal{M}({\bm p})$. If the target bucket has available capacity (occupied by noise points), the DO inserts \(\bm{p}\) directly into the bucket, replacing the noise points. Otherwise, if the bucket is full, the DO creates a new bucket \(\bm{B}\), inserts \(\bm{p}\) into \(\bm{B}\), and appends \(\bm{B}\) to the bucket set.
Subsequently, DO updates the leaf predictor’s error bounds accordingly, encrypts the updated index as $\widetilde{I}^{e}$, and uploads it to the DSP for replacement.

{\bf Deletion.} To delete $\bm p$, the DO marks it as deleted via the bitmap index within its bucket. 
The bucket is removed only when empty, after which the leaf predictor’s error bounds are updated. Finally, the updated index $\widetilde{I}^{e}$, including the bitmap, is uploaded to the DSP.

Notably, frequent insertions and deletions may distort data distribution and degrade query accuracy. When updates exceed a threshold, the DO should rebuild \gls{idx}.

\begin{algorithm}[t]
	\caption{Secure Bucket Prediction}
	\scriptsize
	\label{alg01}
	\LinesNumbered 
	\KwIn{Encrypted query $\llbracket Q\rrbracket$ and \gls{idx} $I^{e}$.}
	\KwOut{Prediction bucket $pb$ or $\mathrm{NULL}$.}
	
	\tcp{{DSP:}}
	$i=0$, $j=0$\;
	\While{${\mathcal{M}}_{(i,j)}^{e} \in I^{e}$ is not a leaf predictor}{
		$j = \widetilde{\mathcal{M}}_{(i,j)}^{e}(\llbracket Q\rrbracket)$\;
		$i=i+1$\;
	}
	$\llbracket v \rrbracket = {\mathcal{M}}_{(i,j)}^{e}(\llbracket Q \rrbracket)$\;
	$\llbracket v^{{\prime}} \rrbracket = \mathrm{SRAND}(1,size({\bm B}),\llbracket v \rrbracket)$\;
	\tcp{\small $r_{1}$ is a random number}
	$\llbracket v \rrbracket = \llbracket v \rrbracket \times {\llbracket r_{1} \rrbracket}$; randomly generate $v^{{\prime}}$\;
	add $\llbracket v \rrbracket$ and $\llbracket v^{{\prime}} \rrbracket$ into $V$; send $V$ to DAP\;
	\tcp{{DAP:}}
	$\widetilde{V} = D_{sk}(\llbracket V \rrbracket)$; send $\widetilde{V} = \pi_{2}(\widetilde{V})$ to DSP\;
	\tcp{{DSP:}}
	\For{$v \in \widetilde{V}$}{
		$v = v - r_{1}$\;
		\For{$E(\bm{p}_{j}) \in \bm{B}_{v}$}{
			\tcp{\small ${r}_{2}$ is a random number}
			$E(\widetilde{\bm{p}}_{j}) = E(\bm{p}_{j}) \times E(r_{2})	$	\;
			add $E(\widetilde{\bm{p}}_{j}) $ into $\widetilde{R}$\;
		}
	}
	send $\widetilde{R} = \pi_{2}(\widetilde{R})$ and $\llbracket Q \rrbracket = \llbracket Q \rrbracket \times E(r_{2})$ to DAP\;
	\tcp{{DAP:}}
	$\tilde{Q} = D_{sk}(\llbracket Q \rrbracket)$\;
	$s = 0$\;
	\For{$\llbracket {\tilde{\bm p}_{j}} \rrbracket \in \widetilde{R}$}{
		${\tilde{\bm p}_{j}} = D_{sk}(\llbracket {\tilde{\bm p}}_{j} \rrbracket)$\;
		\If{${\tilde{\bm p}_{j}} = {\tilde{Q}}$}{
			$s = 1$\;
		}
	}
	send $s$ to DSP\;
	\tcp{{DSP:}}
	\If{$s=0$}{
		\Return $\mathrm{NULL}$\;
	}
	\Return $pb = \llbracket v \rrbracket$\;
\end{algorithm}

\section{Secure Range Queries}

\subsection{Secure Bucket Prediction}
Our \gls{sbp} aims to iteratively select the corresponding next-level predictor while concealing query trajectories through the use of dummy buckets, which achieves indistinguishability in both search patterns and access patterns.

As demonstrated in Algorithm \ref{alg01}, SBP takes an encrypted query $\llbracket Q \rrbracket$ and a secure index $I^{e}$ as inputs. The process initiates with the secure root predictor $\widetilde{\mathcal{M}}^{e}_{(0,0)}$, which recursively invoking its sub-predictors until reaching a secure leaf predictor. Notably, at each level, only one sub-predictor is accessed. 
Let $\widetilde{\mathcal{M}}^{e}_{(i,j)}$ denote the predictor accessed at the $i$-th intermediate level.
The corresponding sub-predictor at layer $i+1$ is denoted by $\widetilde{\mathcal{M}}^{e}{(i+1,j^{\prime})}$, where $j^{\prime}$ is computed as $j^{\prime} = \widetilde{\mathcal{M}}^{e}{(i,j)}(\llbracket Q \rrbracket)$. 
To conceal the specific bucket position, DSP first obtains an encrypted prediction identifier $\llbracket v \rrbracket$ from a leaf predictor ${\mathcal{M}}^{e}{(i,j)}$, and randomly selects an encrypted bucket identifier $\llbracket v^{\prime} \rrbracket$ ($v \neq v^{\prime}$). 
The identifiers $\llbracket v \rrbracket$ and $\llbracket v^{\prime} \rrbracket$ are subsequently obfuscated using a random scalar $r_1 \in \mathbb{Z}_N$, generated by a pseudo-random function $f$.
Thereafter, $\llbracket v \rrbracket$ and $\llbracket v^{{\prime}} \rrbracket$ are embedded into a vector $V$, which is transmitted to the DAP for subsequent processing.

Upon receiving the vector $V$, DAP decrypts $V$ to obtain $\widetilde{V}$, which is then shuffled via a random permutation function $\pi_{2}$ before being sent back to DSP. Subsequently, DSP accesses the corresponding buckets ${\bm B}_{v}$ using the values in $\widetilde{V}$ as indices. 
For each data point $E({\bm p}_{j}) \in {\bm B}_{v}$, a random noise perturbation $r_{2}$ generated by $f$ is applied to obtain the obfuscated point $E(\widetilde{{\bm p}}_{j})$, which is then aggregated into the set $\widetilde{R}$. 
Then, DSP permutes $\widetilde{R}$ using a random permutation function $\pi_{2}$ and sends the permuted set along with the noise-perturbed encrypted query $\llbracket Q \rrbracket$ to DAP. Note that, since the data points and the query are obfuscated using the same $r_2$, their relative positions remain unchanged.

After receiving these values, DAP first decrypts the query $\llbracket Q \rrbracket$ to obtain $\widetilde{Q}$, then initializes a flag $s=0$. For each obfuscated point $\llbracket \widetilde{{\bm p}}_{j} \rrbracket \in \widetilde{R}$, DAP decrypts  $\llbracket \widetilde{{\bm p}}_{j} \rrbracket$ and checks whether $\llbracket {{\bm p}}_{j} \rrbracket$ matches $\widetilde{Q}$. 
If they match, $s$ is set to 1, otherwise $s=0$. 
Thereupon, the flag $s$ is transmitted to DSP, which determines the result accordingly: if $s = 1$, it returns the encrypted prediction index $\llbracket v \rrbracket$, otherwise it returns $\mathrm{NULL}$.

\begin{algorithm}[t]
	\caption{Secure Point Extraction}
		\scriptsize
	\label{alg02}
	\LinesNumbered 
	\KwIn{Encrypted query $\llbracket Q\rrbracket = \langle \llbracket {\bm{q}_{l}}\rrbracket, \llbracket {\bm{q}_{r}}\rrbracket \rangle$ and $I^{e}$.}
	\KwOut{Encrpyted candidate result $R^{{\prime}}$.}
	
	\tcp{{DSP:}}
	\If{$\mathrm{SBP}(\llbracket \bm{{q}_{l}} \rrbracket)$ is $\mathrm{NULL}$}{
		$\llbracket\beta_{l} \rrbracket = \mathrm{SBP}(\llbracket {\bm{q}_{l}} \rrbracket) \times \llbracket {err}_{max} \rrbracket^{-1}$\;
	}\Else{
		$\llbracket\beta_{l} \rrbracket = \mathrm{SBP}(\llbracket {\bm{q}_{l}} \rrbracket)$\;	
	}
	\If{$\mathrm{SBP}(\llbracket \bm{{q}_{r}} \rrbracket)$ is $\mathrm{NULL}$}{
		$\llbracket\beta_{r} \rrbracket = \mathrm{SBP}(\llbracket {\bm{q}_{r}} \rrbracket) \times \llbracket {err}_{max} \rrbracket$\;
	}\Else{
		$\llbracket\beta_{r} \rrbracket = \mathrm{SBP}(\llbracket {\bm{q}_{r}} \rrbracket)$\;
	}
	send $\llbracket \beta_{l} \rrbracket = \llbracket\beta_{l} \rrbracket \times \llbracket r_{1} \rrbracket $ \& $\llbracket \beta_{r} \rrbracket = \llbracket\beta_{r} \rrbracket \times \llbracket r_{1} \rrbracket$ to DAP\;
	\tcp{{DAP:}}
	$\beta_{upp} = D_{sk}(\llbracket\beta_{r} \rrbracket)$; $\beta_{low} = D_{sk}(\llbracket\beta_{l} \rrbracket)$\;
	send $\langle\beta_{low},\beta_{upp}\rangle$ to DSP\;
	\tcp{{DSP:}}
	$\beta_{low} = \beta_{low}-r_{1}$; $\beta_{upp} = \beta_{upp}-r_{1}$\;
	
	\For{$i \in [\beta_{low},\beta_{upp}]$ }{
		\For{$ E({\bm p}_{j}) \in {\bm B}_{i}$}{
			\tcp{\small $ r_{i}$ is a random number}
			$ E(\widetilde{\bm p}_{j}) = E({\bm p}_{j}) \times \llbracket {\bm r}_{i} \rrbracket$;
			add $ E(\widetilde{\bm p}_{j})$ into $\widetilde{\bm B}_{i}$\;
		}
		$E(\widetilde{\bm{mbr}}_{i})= {\bm B}_{i}.E(\bm{mbr}) \times \llbracket {r}_{2} \rrbracket$;
	}	
	
	$\llbracket \widetilde{Q} \rrbracket = \llbracket Q \rrbracket \times \llbracket r_{2} \rrbracket$\; 
	pack $\llbracket \widetilde{Q}\rrbracket$ into $\theta$, $\pi_{2}(E(\widetilde{\bm{mbr}}))$ into $\delta$\; 
	send $\theta$, $\delta$ and $\widetilde{\bm B} = \pi_{2}(\widetilde{\bm B})$ to DAP\; 
	\tcp{{DAP:}}
	unpack $D_{sk}(\theta)$ to $\widetilde{Q}$, $D_{sk}(\delta)$ to $\widetilde{\bm{mbr}}$; 
	
	\For{$i=1$ to $ size(\widetilde{\bm B})$}{
		\If{$\widetilde{\bm{mbr}}_{i}$ intersects $\widetilde{Q}$}{
			initialize a vector ${\bm u}$ with $E(0)$\;
			${\bm u}_{i} = E(1)$;
			add ${\bm u}$ into $U$\;
			\For{$E(\widetilde{\bm p}_{j}) \in \widetilde{\bm B}_{i}$}{
				add $E(\widetilde{\bm p}_{j})$ into $\widetilde{R}^{{\prime}}_{i}$\;
			}
		}

	}
	send $U$ and $\widetilde{R}^{{\prime}}$ to DSP\;
	\tcp{{DSP:}}
	\For{$i = 1$ to $|U|$}{
		$ U^{{\prime}}_{i} = \pi_{2}^{-1}(U_{i})$;	
		$\Theta_{i} = \prod_{j=1}^n (U_{i}[j])^{{\bm r}_{i}}$\;
		\For{$E(\widetilde{\bm p}_{j}) \in \widetilde{R}^{{\prime}}_{i}$}{
			$E({\bm p}_{j}) = E(\widetilde{\bm p}_{j}) \times \Theta_{i}^{-1}$\;
			add $E({\bm p}_{j})$ into $R^{{\prime}}$\;
		}
		
	}
	\Return $R^{{\prime}}$\;	
\end{algorithm}

\subsection{Secure Point Extraction} 

Our \gls{spe} is designed to securely retrieve encrypted points within target buckets without compromising the privacy of datasets, queries, results, and access pattern. 
A straightforward solution is to design a secure intersection protocol that determines whether the query intersects with points in the bucket, and then extracts the matching points using a secure multi-party (SM) protocol. However, this approach is computationally expensive. To address this, we propose an efficient \gls{spe} protocol that significantly reduces the  computational overhead, particularly the decryption cost.

As demonstrated in Algorithm \ref{alg02}, \gls{spe} takes an encrypted range query $\llbracket Q \rrbracket = (\llbracket {\bm q_{l}} \rrbracket,\llbracket {\bm q_{r}} \rrbracket)$ and a \gls{idx} $I^{e}$ as inputs, and outputs an encrypted candidate point set $R^{{\prime}}$. Initially, DSP inputs $\llbracket \bm q_{l} \rrbracket$ and $\llbracket \bm q_{r} \rrbracket$ into the SBP protocol to obtain the encrypted bucket positions $\llbracket \beta_{l} \rrbracket$ and $\llbracket \beta_{r} \rrbracket$. Notably, if SBP fails to locate $\llbracket \bm q_{l} \rrbracket$ or $\llbracket \bm q_{r} \rrbracket$, the output of the leaf predictor is adjusted using an error tolerance, i.e., $\llbracket \beta_{l} \rrbracket = \mathrm{SBP}(\llbracket {\bm{q}_{l}} \rrbracket) \times \llbracket err_{max} \rrbracket^{-1}$ and $\llbracket \beta_{r} \rrbracket = \mathrm{SBP}(\llbracket {\bm{q}_{r}} \rrbracket) \times \llbracket err_{max} \rrbracket$. Although this adjustment introduces minor overhead, the injected error serves to obscure the true bucket positions, thereby preventing DSP from inferring precise query locations. Additionally, a random noise term $r_{1} \in \mathbb{Z}_{N}$ is homomorphically applied to $\llbracket \beta_{l} \rrbracket$ and $\llbracket \beta_{r} \rrbracket$ to further protect the query range from DAP.

After that, DSP and DAP communicate to obtain the decrypted bucket positions $\beta_{low}$ and $\beta_{upp}$ from $\llbracket \beta_{l} \rrbracket$ and $\llbracket \beta_{r} \rrbracket$. After eliminating the noise $r_{1}$, DSP then traverses the buckets within the range $[\beta_{low},\beta_{upp}]$, where each data point $E({\bm p}_{j}) \in B_{i}$ is perturbed with a random noise ${\bm r}_{i}$ and then aggregated into a new bucket $\widetilde{\bm B}_{i}$. Subsequently, the encrypted query $E(Q)$ and the encrypted MBR $E({\bm {mbr}})$ of each bucket ${\bm B}_{i}$ are obfuscated using a noise $r_{2}$, resulting in $\llbracket \widetilde{Q} \rrbracket$ and $E(\widetilde{\bm{mbr}}_{i})$. The use of the same noise $r_{2}$ for both $Q$ and ${\bm {mbr}}$ ensures a consistent window offset for comparison operations. After obfuscation, DSP applies a random permutation function $\pi_{2}$ to shuffle $E(\widetilde{\bm{mbr}})$ and then packs them into $\delta$, while the obfuscated query $\llbracket \widetilde{Q} \rrbracket$ is packed into $\theta$.
Next, DSP sends $\theta$, $\delta$, and the permuted set of obfuscated buckets $\widetilde{\bm B}$ (also shuffled using $\pi_{2}$) to the DAP.

Upon receiving these values, the DAP first decrypts $\theta$ and $\delta$, and unpacks them to $\widetilde{Q}$ and $\widetilde{\bm{mbr}}$, respectively. Since both have been obfuscated using the same noise, it is straightforward to determine whether $\widetilde{Q}$ intersects with each $\widetilde{\bm{mbr}}$ in the plaintext domain. If an intersection is detected,  DAP initializes an $n$-dimensional vector ${\bm u}$ with all elements set to 0, then sets ${\bm u}_{j} = E(1)$ to indicate the position of the corresponding bucket. This vector ${\bm u}$ is added to the set $U$. Meanwhile, the data points within the intersecting bucket are aggregated into a new set $\widetilde{R}^{{\prime}}_{i}$, which contains only the encrypted points from $\widetilde{\bm B}_{i}$. Once all relevant buckets have been processed, the DAP sends both $U$ and $\widetilde{R}^{{\prime}}$ to the DSP.

Recall that each encrypted point was perturbed with a noise $r_{i}$, which DSP need subsequently remove. The DSP first applies the inverse permutation $\pi_{2}^{-1}$ to each element $U_{i}$ in $U$, since the bucket set $
\widetilde{\bm B}$ was permuted by $\pi_{2}$ prior to transmission to DAP. Each element $U_{i}$ is a vector in which only the element corresponding to the target bucket is set to $E(1)$, while all other elements are set to $E(0)$. Consequentially, DSP computes a denoising factor for each target bucket as $\Theta_{i} = \prod_{j=1}^n (U_{i}[j])^{{\bm r}_{i}}$. Finally, DSP applies $\Theta_{i}$ to each encrypted point in the corresponding bucket $\widetilde{R}_{i}^{{\prime}}$ to remove the noise and recover $E({\bm p}_{j})$, which are then aggregated into the final candidate result set.

\begin{table*}[t]
	\centering
	\renewcommand\arraystretch{1.11}
	\caption{Example of FSRQ ($\sigma = 2,\lambda  = 5$).}
	\vspace{-3mm}
	\label{tab03}
	\begin{tabular}{|c|c|c|c|c|c|c|c|c|c|}
		\hline
		$V$ &
		$T$ &
		\rule{0pt}{3ex}$\widetilde{T}$&
		$\bm{r}$ &
		$\nu$ &
		$\Upsilon$ &
		$\Gamma$ &
		$\widetilde{R}$ &
		$R$ \\ \hline
		$(1,0,1,0,0)$ &
		\begin{tabular}[c]{@{}c@{}}$\{{\bm{p}}_{1}, {\bm{p}}_{2}, {\bm{p}}_{3},{\bm{p}}_{4},{\bm{p}}_{5}\}$\end{tabular} &
		\begin{tabular}[c]{@{}c@{}}$\{\widetilde{{\bm{p}}}_{5},\widetilde{{\bm{p}}}_{4}, \widetilde{{\bm{p}}}_{3},\widetilde{{\bm{p}}}_{2},\widetilde{{\bm{p}}}_{1}\}$\end{tabular} &
		\begin{tabular}[c]{@{}c@{}}${\bm{r}}_{1}=1$, ${\bm{r}}_{2}=9$\\ ${\bm{r}}_{3}=2$, ${\bm{r}}_{4}=7$ \\${\bm{r}}_{5}=8$\end{tabular} &
		17 &
		\begin{tabular}[c]{@{}c@{}}$\Upsilon_{1}=(0,0,1,0,0)$\\ $\Upsilon_{2}=(0,0,0,0,1)$\end{tabular} &
		\begin{tabular}[c]{@{}c@{}}$\Gamma_{1}=2$\\ $\Gamma_{2}=1$\end{tabular}&
		\begin{tabular}[c]{@{}c@{}}$\{\widetilde{{\bm{p}}}_{3},\widetilde{{\bm{p}}}_{1}\}$\end{tabular}&
		\begin{tabular}[c]{@{}c@{}}$\{{\bm{p}}_{3},{\bm{p}}_{1}\}$\end{tabular} \\ \hline
	\end{tabular}
	\vspace{-3mm}
\end{table*}

\begin{algorithm}[t]
	\caption{\gls{idx}-Based Range Queries}
		\scriptsize
	\label{alg03}
	\LinesNumbered 
	\KwIn{Encrypted range query $\llbracket Q\rrbracket$ and \gls{idx} $I^{e}$.}
	\KwOut{Encrpyted result $R^{e}$.}
	
	\tcp{{DSP:}}
	
	$T = \mathrm{\gls{spe}}(\llbracket Q \rrbracket,I^{e})$\;
	\For{$E(\bm{p}_{i}) \in T$}{
		$E(V_{i}) = \mathrm{SQP}(E({\bm p}_{i}),\llbracket Q\rrbracket)$\;
		\tcp{\small $r_{i}$ is a random number}
		$E(\widetilde{{\bm p}}_{i}) = E({\bm p}_{i}) \times \llbracket {\bm r}_{i} \rrbracket$\;
		add $E(\widetilde{{\bm p}}_{i})$ into $\widetilde{T}$\;
	}
	pack $ E(V) = \pi_{2}(E(V))$ into $\nu$\;
	
	send $\widetilde{T} = \pi_{2}(\widetilde{T})$ and $\nu$ to DAP\;
	\tcp{{DAP:}}
	$\nu^{{\prime}}= D_{sk}(\nu)$;
	unpack $\nu^{{\prime}}$ to $\widetilde{V}$\;
	\For{$i = 1$ to $|\widetilde{V}|$}{
		\If{$\widetilde{V}_{i} = 1$}{
			initialize a vactor ${\bm \upsilon}$ with $E(0)$\;
			${\bm \upsilon}_{i} = E(1)$\;
			add ${\bm \upsilon}$ into $\Upsilon$\;
			add $\widetilde{T}_{i}$ into $\widetilde{R}$\;	
		}
		
	}
	send $\Upsilon$ and $\widetilde{R}$ to DSP\;
	\tcp{{DSP:}}
	\For{$i=1$ to $|\Upsilon|$}{
		$\Upsilon^{{\prime}}_{i} = \pi_{2}^{-1}(\Upsilon_{i})$\;
		$\Gamma_{i} = \prod_{j=1}^{n} (\Upsilon^{{\prime}}_{i}[j])^{\bm{r}_i} $\;	
		$ E(\bm{p}_{i})=\widetilde{R}_{i} \times \Gamma^{-1} $\;
		add $E(\bm{p}_{i})$ into $R^{e}$\;
	}
	\Return $R^{e}$\;	
\end{algorithm}

\subsection{Secure Range Queries with \gls{idx}}
Our \gls{RQ} is designed to support efficient and secure range queries using \gls{spe} in conjunction with a \gls{idx}.
While \gls{spe} significantly reduces the number of points requiring secure scanning through prediction, efficiently filtering the points within the predicted range remains a challenge. To address this, we introduce the \gls{RQ} protocol, which incorporates dynamic noise perturbation and secure permutation primitives to reduce computational overhead while preserving access pattern privacy, thus achieving an effective balance between security and performance in encrypted query processing.

As illustrated in Algorithm \ref{alg03}, \gls{RQ} processes an encrypted range query $\llbracket Q \rrbracket$ over a \gls{idx} $I^{e}$. Initially, DSP invokes the \gls{spe} protocol to securely derive a candidate point set $T$ based on $\llbracket Q \rrbracket$. For each point $E({\bm p}_{j}) \in T$, DSP applies a noise ${\bm r}_{i} \in \mathbb{Z}_{N}$ to generate an obfuscated point $E(\widetilde{\bm{p}}_{j}) = E({\bm p}_{j}) \times \llbracket {\bm {r}_{i}} \rrbracket$, which is then added to the set $\widetilde{T}$.
Meanwhile, DSP constructs a marking vector $V$ where each entry $V_{i}$ is computed using the function $\mathrm{SQP}(E({\bm p}_{j}),\llbracket Q \rrbracket)$. The $\mathrm{SQP}$ function \cite{wang2022efficient} securely determines whether $E({\bm p}_{j})$ is contained in the query range $\llbracket Q \rrbracket$, returning $E(1)$ if it does and $E(0)$ otherwise.
Subsequently, DSP packs $E(V)$, which is scrambled by the random permutation function $\pi_{2}$, into $\nu$, and then transmits $\nu$ along with the permuted candidate set $\pi_{2}(\widetilde{T})$ to DAP. This ensures that the permutation applied to $E(V)$ is consistently aligned with that of $\widetilde{T}$.

After receiving these values, DAP only decrypts $\nu$ to $\nu^{{\prime}}$ and unpacks it to $\widetilde{V}$, where each entry $\widetilde{V}_{i} = 1$ means the corresponding point falls within the query $\llbracket Q\rrbracket$. Specifically, DAP checks all elements in $\widetilde{V}$, and for each $\widetilde{V}_{i} = 1$, the corresponding point is added into the intermediate result set $\widetilde{R}$. Meanwhile, DAP initializes an $n$-dimensional binary vector ${\bm v}$ with $E(0)$ and marks the corresponding position by setting ${\bm v}_{i} = E(1)$. The marked vector ${\bm v}$ is then added to the set $\Upsilon$. Thereafter, DAP transmits both $\Upsilon$ and $\widetilde{R}$ to DSP for subsequent processing.

To obtain the final result while protecting the access pattern privacy, DSP needs to remove the noise from the corresponding points. To begin with, DSP applies the inverse permutation $\pi_{2}^{-1}$ to each element $\Upsilon_{i}$ to obtain $\Upsilon_{i}^{{\prime}}$, and then computes the corresponding noise aggregation factor as $\Gamma_{i} = \prod_{j=1}^{n} (\Upsilon^{{\prime}}_{i}[j])^{\bm{r}_i}$. Finally, the DSP removes the noise by computing $E(\bm{p}_{i})=\widetilde{R}_{i} \times \Gamma^{-1}$, thereby recovering the encrypted result point, which is added to the result set $R^{e}$.

{\bf Example.} 
Table \ref{tab03} illustrates how \gls{RQ} securely processes range queries while preserving data confidentiality. We assume the candidate point set returned by the \gls{spe} protocol is $T = \{{\bm p}_{1},{\bm p}_{2},{\bm p}_{3},{\bm p}_{4},{\bm p}_{5}\}$, where ${\bm p}_{1}$ and ${\bm p}_{3}$ satisfy the query constraints. The DSP scans all points in \( T \), constructs a marking vector \( V = (1,0,1,0,0) \), and perturbs each \( {\bm p}_i \) with noise \( {\bm r}_i \) to form \( \widetilde{T} \). A permutation \( \pi_2 = (5,4,3,2,1) \) is then applied to both \( \widetilde{T} \) and \( V \), yielding \( \pi_2(V) = (0,0,1,0,1) \), which is compacted into \( \nu = 17 \) via $\left\langle x_1 \vert \dots \vert x_\lambda \right\rangle = \sum_{i=1}^\lambda x_i 2^{\sigma (\lambda - i)}$.
Next, DAP decrypts \( \nu \), unpacks it to obtain \( \widetilde{V} \), and identifies target positions to construct corresponding vectors \( \Upsilon \), i.e., \( \Upsilon_1 = (0,0,1,0,0) \), \( \Upsilon_2 = (0,0,0,0,1) \). Corresponding encrypted points are aggregated into \( \widetilde{R} \), followed by noise correction using inverse permutation \( \pi_2^{-1} \) and noise factor \( \Gamma \), enabling recovery of the original result set. Note that all values in Tables~\ref{tab03} are shown in encrypted form.

\section{Complexity and Security Analysis}\label{analy}

\subsection{Complexity Analysis}
We analyze the complexity of our \gls{RQ} protocol from two perspectives: \textit{communication complexity} and \textit{computation complexity}. Here, $\|\mathbb{N}\|$ denotes the bit-length of a ciphertext produced by the corresponding encryption scheme.

The computational complexity analysis of \gls{RQ} is as follows.
Since \gls{idx} organizes secure predictors in a tree structure, the overall \gls{RQ} achieves logarithmic-time execution with $O(d\log n)$ computational complexity. Furthermore, by organizing encrypted data into a hierarchical index structure, \gls{RQ} achieves a computational complexity of $O(nd\log n)$ and incurs a cost of $O(d)$ for trapdoor generation. Additionally, \gls{RQ} supports dynamic index updates, which incur an overhead of $O(d\log n )$.

In terms of communication complexity, \gls{RQ} requires an overall query communication cost of $O(d\log n \cdot \|\mathbb{N}\|)$ bits. Since secure indexes need to be uploaded to the cloud, the communication overhead incurred in this process is $O(nd\log n \cdot \|\mathbb{N}\|)$. Trapdoor generation incurs a communication overhead of $O(d \cdot \|\mathbb{N}\|)$, as the client transmits encrypted range bounds for each dimension to the cloud for every query. Finally, dynamic updates incur a communication overhead of $O(d \cdot \|\mathbb{N}\|)$.

\subsection{Security Analysis}\label{SecAna}
Recall that Section~\ref{SecurityModel} introduces leakage functions, where $\mathcal{L}_{\text{Build}}$, $\mathcal{L}_{\text{Update}}$, and $\mathcal{L}_{\text{Query}}$ are now detailed.
\begin{itemize}
	\item $\mathcal{L}_{\text{Build}}(\llbracket P \rrbracket) = (n, d, \Psi)$, where $n$ is the number of points, $d$ is the number of dimensions, and $\Psi$ is the sorted order of encoded points. Importantly, $\Psi$ does not reveal attribute-specific values.
	
	\item $\mathcal{L}_{\text{Update}}(I^{e}) = (pos, bid, \varpi)$, where $pos$ indicates the index insertion position, $bid$ denotes the bucket identifier, and $\varpi$ specifies whether the operation is an insertion or deletion.
	
	\item $\mathcal{L}_{\text{Query}}(I^{e}, \llbracket Q\rrbracket) = (\beta_{\text{low}}, \beta_{\text{upp}}, |R^e|)$, where $\beta_{\text{low}}$ and $\beta_{\text{upp}}$ represent the upper and lower bounds of the scan, and $|R^e|$ denotes the size of the result set, i.e., volume pattern.
	
\end{itemize}

Then, we analyze the security of our \gls{RQ} using the framework of the simulation paradigm. 


\begin{theorem}
	\gls{RQ} is $(\mathcal{L}_{\text{Build}}, \mathcal{L}_{\text{Update}}, \mathcal{L}_{\text{Query}})$-secure under the assumptions that $\pi$ and $f$ are pseudo-random and that the Paillier cryptosystem is semantically secure.
\end{theorem}

\begin{proof}
	We construct a PPT simulator $\mathcal{S}$ such that for any PPT adversary $\mathcal{A}$, the simulated and real executions are computationally indistinguishable.
	
	In the game $\textsf{Real}_{\mathcal{A}}$, $\mathcal{A}$ observes the encrypted index $I^{e}$ and trapdoors $E(Q_i)$ and interacts with the protocol to obtain results $R^e$.
	
	In the game $\textsf{Ideal}_{\mathcal{A},\mathcal{S}}$, given $\mathcal{L}_{\text{Build}}$, $\mathcal{S}$ generates a dataset $P^* = \{\bm{p}_1^{*}, ..., \bm{p}_n^{*}\}$ with random dummy values and encoding values. Next, $\mathcal{S}$ builds a simulated index $I^{*} \leftarrow \textsf{BuildIndex}(P^{*}, pk)$, which is computationally indistinguishable from the real index due to the semantic security of Paillier.
	Given $\mathcal{L}_{\text{Update}}(I^{*})$, $\mathcal{S}$ generates a random point to simulate insertion or deletion at the specified bucket and position.
	For each query, $\mathcal{S}$ constructs $Q_i^{*}$ such that its scan range corresponds to the bounds $\beta_{\text{low}}$ and $\beta_{\text{upp}}$. $\mathcal{S}$ then computes the encrypted trapdoor $E(Q_i^{*})$ using the public key. By the semantic security of Paillier, $E(Q_i^{*})$ is indistinguishable from $E(Q_i)$. Next, $\mathcal{S}$ executes the query over the simulated index $I^{*}$ to obtain a simulated encrypted result set $R^{*}$.
	
	Since the results, points, bucket identifiers and predicted point positions are encrypted using the semantically secure Paillier cryptosystem, $\mathcal{A}$ cannot distinguish between the real and simulated views. Besides, due to the semantic security of Paillier and the pseudo-randomness of $f$, $\mathcal{A}$ is unable to infer the access path or link queries to the corresponding data points, thereby preserving access pattern privacy.
	
	Hence, we conclude that the views of $\mathcal{A}$ in the $\textsf{Real}_{\mathcal{A}}$ and $\textsf{Ideal}_{\mathcal{A},\mathcal{S}}$ are indistinguishable, that is,  
	\begin{equation*}
		\left| \Pr[\textsf{Real}_{\mathcal{A}}(\lambda)=1] - \Pr[\textsf{Ideal}_{\mathcal{A},\mathcal{S}}(\lambda)=1] \right| \leq \text{negl}(\lambda).
	\end{equation*}
\end{proof}

\begin{table}[t]
	\centering
	\renewcommand\arraystretch{1.11}
	\caption{Parameter settings.}
	\vspace{-3mm}
	\label{tab04}
	\begin{tabular}{|c|c|}
		\hline
		Parameter Names          & Value                                 \\ \hline
		Size of dataset $n$        & ${\bm {20000}}$ 40000 60000 80000 100,000       \\ \hline
		Key length $K$             & ${\bm {1024}}$ 2048 3072 4096                   \\ \hline
		Bucket capacity $b$        & 4 ${\bm 8}$ 16 32                             \\ \hline
		Predicted point threshold $m$  & ${\bm {300}}$ 500 700 800 1100                  \\ \hline
		Range query size         & ${\bm{ 0.0025\%}}$ 0.005\% 0.01\% 0.02\% 0.04\% \\ \hline
		Range query aspect ratio& ${\bm {0.25}}$ 0.5 1 2 4                        \\ \hline
		Insertion rate        & 2\% 4\% 6\% 8\% 10\%                  \\ \hline
	\end{tabular}
\end{table}

\section{EXPERIMENT}
\subsection{Experiment Setup}
The experiments were implemented in C++ on a PC equipped with an Intel i5 CPU @ 3.50 GHz, 16 GB RAM, and running Ubuntu 20.04. We utilized the C++ API of PyTorch 1.4 to implement CPU-based learned indexes. The DSP, DAP, data owner, and query user were deployed as independent processes. Unless stated otherwise, the reported time reflect the total computation and communication time incurred by both DSP and DAP.

{\bf Datasets.} Following previous work \cite{qi2020effectively}, we generated three synthetic datasets: UNI, NOR, and SKE, each containing up to 100,000 samples. The UNI and NOR datasets follow uniform and normal distributions, respectively. The SKE dataset is constructed by exponentiating the last dimension of uniformly distributed points to introduce skewness. Additionally, we used two real-world datasets: (1) CAR, which contains over 2 million 2D points representing road networks in California\footnote{http://snap.stanford.edu/data/roadNet-CA.html}; and (2) GOW, which includes coordinate samples from user check-ins location\footnote{http://snap.stanford.edu/data/loc-gowalla totalCheckins.txt.gz}.

{\bf Parameters.} \gls{RQ} was evaluated under the following the following settings: (1) Dataset sizes from 20,000 to 100,000 samples; (2) Paillier key lengths from 1024 to 4096 bits; (3) Bucket capacities between 4 and 32; (4) Predicted point threshold (i.e., maximum number of points covered by each predictor) ranging from 300 to 1100; (5) Range query size e from 0.0025\% to 0.04\% of the data space; (6) Range query aspect ratio from 0.25 to 4; (7) Insertion rate from 2\% to 10\% of the dataset. The detailed parameter setting is shown in Table~\ref{tab04} (default values are indicated in bold).

We compared our \gls{RQ} against existing schemes. TRQED$^{+}$ employs cryptographic methods to achieve sub-linear query time while preserving the privacy of datasets, queries, results and access patterns. SRQ$_b$ uses encrypted R-trees with secure node and point intersection protocols but does not protect access or search patterns.

\subsection{\gls{idx} Construction}
{\bf Index construction time.}
As shown in Table~\ref{tab05}, the index construction time of \gls{idx} grows approximately linearly with $n$ across all five datasets. Real-world datasets consistently incur higher construction costs than synthetic ones due to their more skewed data distributions. Specifically, CAR incurs the highest overhead, followed by GOW, with Uniform, Normal, and Skewed in descending order.
Additionally, construction cost also includes node encryption, i.e., mainly encrypting neural network parameters and data points, which increases with tree height as node count grows exponentially.

\begin{table}[t]
	\centering
	\renewcommand\arraystretch{1.11}
	\caption{Index construction time (s).}
	\vspace{-3mm}
	\label{tab05}
	\begin{tabular}{|c|c|c|c|c|c|}
		\hline
		n ($\times 10^{4}$)/Datasets & UNI    & NOR    & SKE    & CAR    & GOW    \\ \hline
		2                 & 27.65  & 11.93  & 25.24  & 38.23  & 38.78  \\ \hline
		4                 & 76.59  & 61.95  & 55.87  & 105.5  & 87.59  \\ \hline
		6                 & 104.92 & 84     & 76.32  & 198.44 & 126.85 \\ \hline
		8                 & 116.52 & 123.04 & 89.05  & 211.84 & 144.08 \\ \hline
		10                & 125.62 & 128.8  & 106.34 & 308.4  & 186.83 \\ \hline
	\end{tabular}
\end{table}
\begin{table}[t]
	\centering
	\vspace{-3mm}
	\renewcommand\arraystretch{1.11}
	\caption{Index storage overhead (MB).}
	\vspace{-3mm}
	\label{tab06}
	\begin{tabular}{|c|*{5}{>{\centering\arraybackslash}m{0.75cm}|}}
		\hline
		n ($\times 10^{4}$)/Datasets & UNI    & NOR    & SKE    & CAR   & GOW   \\ \hline
		2                 & 21.44  & 15.28  & 22.51  &  16.48 &  18.24 \\ \hline
		4                 & 56.95  & 50     & 46.47  &  32.47 &  32    \\ \hline
		6                 & 79.87  & 71     & 68.56  &  47.58 &  45.26 \\ \hline
		8                 & 101.05 & 95.93  & 82.97  &  58.85 &  58.19 \\ \hline
		10                & 119.56 & 113.08 & 100.53 &  73.74 &  82.17 \\ \hline
	\end{tabular}
\end{table}

{\bf Index storage overhead.} 
Table~\ref{tab06} reports the storage overhead across datasets of varying $n$. Overall, storage overhead  remains relatively stable due to fixed key lengths and ciphertext sizes. 
Additionally, the storage overhead for all datasets grows with increasing $n$.

{\bf Impact of other parameters.} 
Table~\ref{tab07} shows the impact of \( m \) on index construction using the Uniform dataset with 20{,}000 samples. As \( m \) increases, both construction time and storage overhead of \gls{idx} decrease and eventually converge, since fewer predictors lead to a shallower index tree, stabilizing at two levels. While optimal \( m \) values vary across datasets with different \( n \), the construction trends remain consistent. Although larger \( m \) reduces construction cost, it does not always yield better query performance. Thus, \( m \) is selected by balancing indexing efficiency and query effectiveness.

\begin{figure}[t]
	\begin{minipage}{\linewidth}
			\begin{minipage}[h]{0.5\linewidth}
			\centering
			\subfigure[Index construction time]{\includegraphics[width=1.79in]{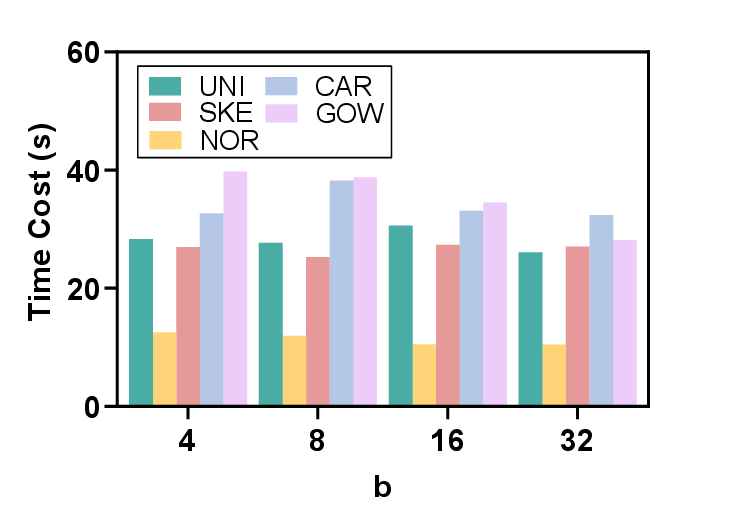}\label{fig:01_a}}\hspace{1mm}
		\end{minipage}%
		\begin{minipage}[h]{0.5\linewidth}
			\centering
			\subfigure[Index storage cost]{\includegraphics[width=1.79in]{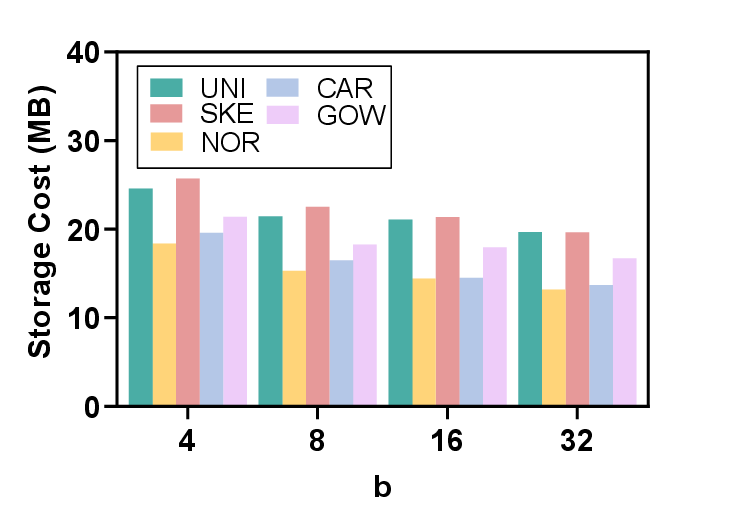}\label{fig:01_b}}
		\end{minipage}%
		\vspace{-3mm}
		\caption{Impact of $b$ on index construction.}
		\label{fig01}
	\end{minipage}
	\vspace{-5mm}
		\begin{minipage}{\linewidth}
			\begin{minipage}[h]{0.5\linewidth}
			\centering
			\subfigure[Index construction time]{\includegraphics[width=1.79in]{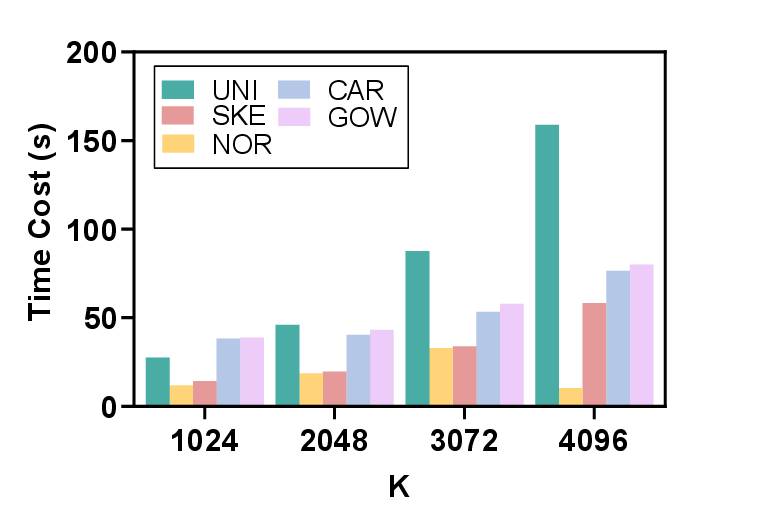}\label{fig:02_a}}\hspace{1mm}
		\end{minipage}%
		\begin{minipage}[h]{0.5\linewidth}
			\centering
			\subfigure[Index storage cost]{\includegraphics[width=1.79in]{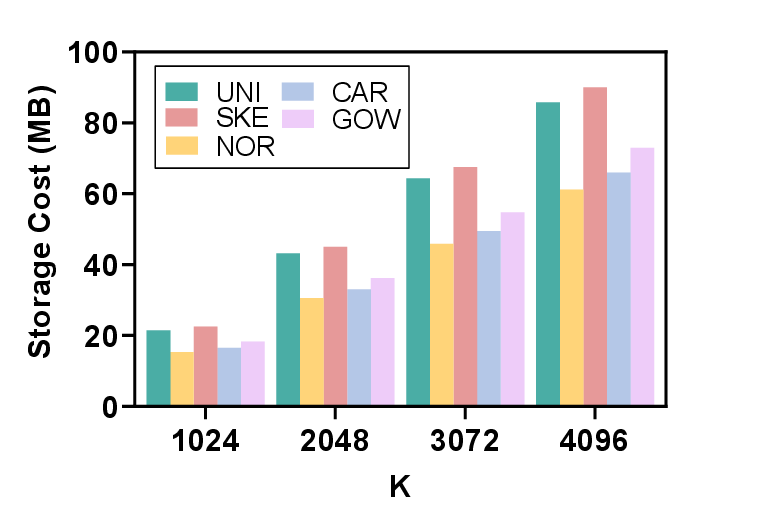}\label{fig:02_b}}
		\end{minipage}%
	\vspace{-3mm}
		\caption{Impact of $K$ on index construction.}
		\label{fig02}
	\end{minipage}
	
	\vspace{5mm}
		\begin{minipage}{\linewidth}
		\begin{minipage}[h]{0.5\linewidth}
			\centering
			\subfigure[Index construction time]{\includegraphics[width=1.79in]{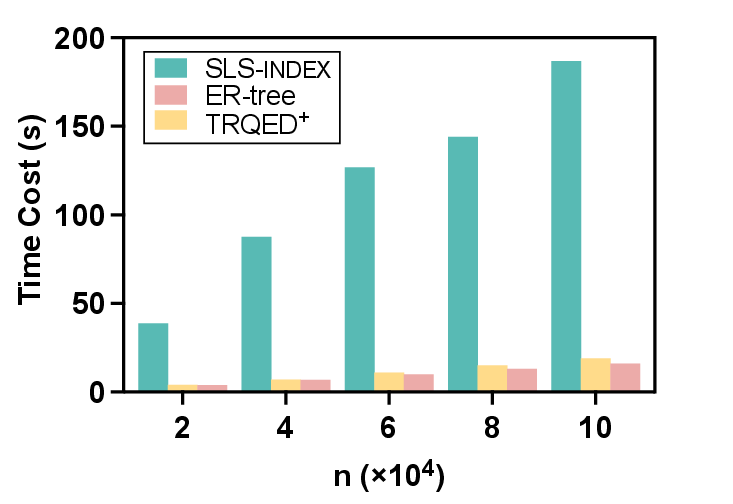}\label{fig:03_a}}\hspace{1mm}
		\end{minipage}%
		\begin{minipage}[h]{0.5\linewidth}
			\centering
			\subfigure[Index storage cost]{\includegraphics[width=1.79in]{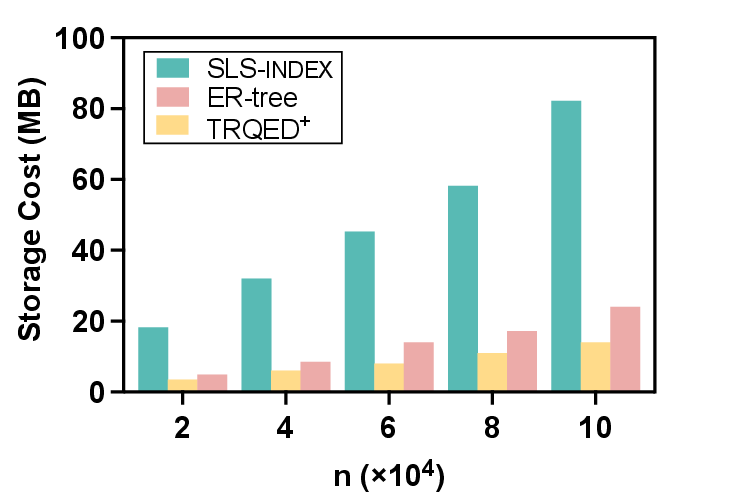}\label{fig:03_b}}
		\end{minipage}%
		\vspace{-3mm}
		\caption{Index construction (\gls{idx} vs. TRQED$^{+}$ vs. ER-tree).}
		\label{fig03}
	\end{minipage}

\end{figure}

As shown in Fig.~\ref{fig01}, we evaluate the effect of bucket capacity \( b \) on index construction. Fig.~\ref{fig01}(a) shows that construction time remains relatively stable across varying \( b \), with only minor fluctuations. In contrast, Fig.~\ref{fig01}(b) indicates that storage overhead decreases as \( b \) increases, since a larger \( b \) reduces the number of buckets without affecting the tree structure.

As shown in Fig.~\ref{fig02}, we present the impact of key length $K$ on index construction.
Fig.~\ref{fig02}(a) shows that construction time overhead increases exponentially with \( K \), indicating that stronger security (larger \( K \)) comes at the cost of higher computational cost.
As shown in Fig.~\ref{fig02}(b), index storage overhead increases linearly with $K$.

\begin{table}[t]
	\centering
	\renewcommand\arraystretch{1.11}
	\caption{Impact of $m$ on index construction.}
	\vspace{-3mm}
	\label{tab07}
	\begin{tabular}{|c|*{5}{>{\centering\arraybackslash}m{0.55cm}|}}
		\hline
		$m$                     & 300   & 500   & 700   & 900   & 1100  \\ \hline
		Construction time (s) & 27.65 & 18.56 & 12.38 & 11.26 & 11.39 \\ \hline
		Storage overhead (MB)     & 21.44 & 20.56 & 18.43 & 17.15 & 16.9  \\ \hline
	\end{tabular}
\end{table}
\begin{table}[t]
	\centering
	\vspace{-3mm}
	\caption{Impact of $m$ on query efficiency.}
	\vspace{-2.5mm}
	\label{tab08}
	\begin{minipage}[t]{1\linewidth}
		\centering
		\subfigure[UNI]{
			\begin{tabular}{|c|*{5}{>{\centering\arraybackslash}m{0.75cm}|}}
				\hline
				$m$            & 200  & 300  & 400  & 500  & 600  \\ \hline
				Query time (s) & 0.55 & 0.49 & 0.56 & 0.59 & 0.97 \\ \hline
			\end{tabular}
			\label{tab07:01}
		}
	\end{minipage}
	\begin{minipage}[t]{1\linewidth}
		\centering
		\vspace{-0.5mm}
		\subfigure[NOR]{	\vspace{-2.5mm}
			\begin{tabular}{|c|*{5}{>{\centering\arraybackslash}m{0.75cm}|}}
				\hline
				$m$            & 600  & 700  & 800  & 900  & 1000 \\ \hline
				query time (s) & 0.99 & 0.95 & 0.95 & 1.02 & 1.03 \\ \hline
			\end{tabular}
			\label{tab07:02}
		}
	\end{minipage}
	\begin{minipage}[t]{1\linewidth}
		\centering
		\vspace{-0.5mm}
		\subfigure[SKE]{
			\begin{tabular}{|c|*{5}{>{\centering\arraybackslash}m{0.75cm}|}}
				\hline
				$m$            & 400  & 500  & 600  & 700  & 800  \\ \hline
				Query time (s) & 1.13 & 1.12 & 1.06 & 1.07 & 1.13 \\ \hline
			\end{tabular}
			\label{tab07:03}
		}
	\end{minipage}
	\begin{minipage}[t]{1\linewidth}
		\centering
		\vspace{-0.5mm}
		\subfigure[CAR]{
			\begin{tabular}{|c|*{5}{>{\centering\arraybackslash}m{0.75cm}|}}
				\hline
				$m$           & 800  & 900  & 1000 & 1100 & 1200 \\ \hline
				Query time (s) & 2.55 & 2.33 & 2.15 & 2.27 & 2.59 \\ \hline
			\end{tabular}
			\label{tab07:04}
		}
	\end{minipage}
	\begin{minipage}[t]{1\linewidth}
		\centering
		\vspace{-0.5mm}
		\subfigure[GOW]{	
			\begin{tabular}{|c|*{5}{>{\centering\arraybackslash}m{0.75cm}|}}
				\hline
				$m$            & 2000 & 2100 & 2200 & 2300 & 2400 \\ \hline
				Query time (s) & 1.38 & 1.05 & 0.98 & 1.05 & 1.08 \\ \hline
			\end{tabular}
			\label{tab07:05}
		}
	\end{minipage}
\end{table}

{\bf Comparing with prior secure indexes.} Fig.~\ref{fig03} compares the index construction time and storage overhead of \gls{idx}, ER-tree, and TRQED$^{+}$ on the GOW dataset. \gls{idx} incurs higher construction time due to the additional cost of model training and encryption of neural network parameters. In contrast, ER-tree and TRQED$^{+}$ are based on plaintext R-trees with encrypted nodes, but TRQED$^{+}$ involves costlier matrix operations for assertion encryption, whereas ER-tree directly encrypts raw node parameters. \gls{idx} also has the highest storage overhead, stemming from the encryption of both model and MBR parameters. As $n$ increases, the exponential growth in nodes further amplifies this overhead.

\begin{figure}[t]
	\begin{minipage}[h]{0.5\linewidth}
		\centering
		\subfigure[Insertion time]{\includegraphics[width=1.79in]{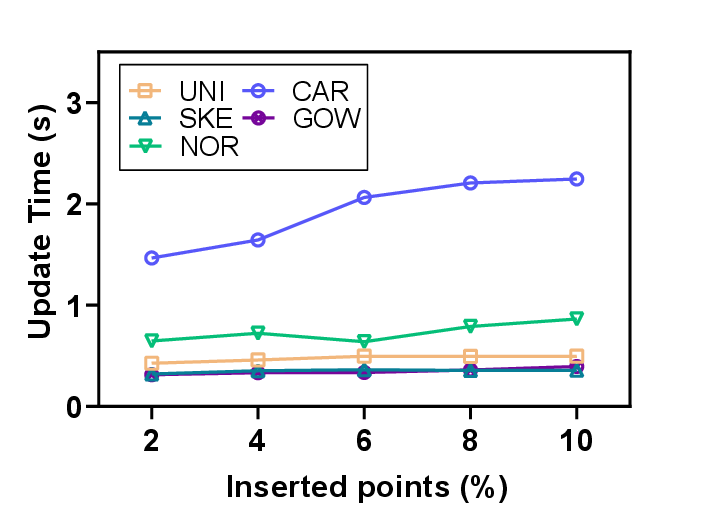}\label{fig:04_a}}\hspace{1mm}
	\end{minipage}%
	\begin{minipage}[h]{0.5\linewidth}
		\centering
		\subfigure[Query time]{\includegraphics[width=1.79in]{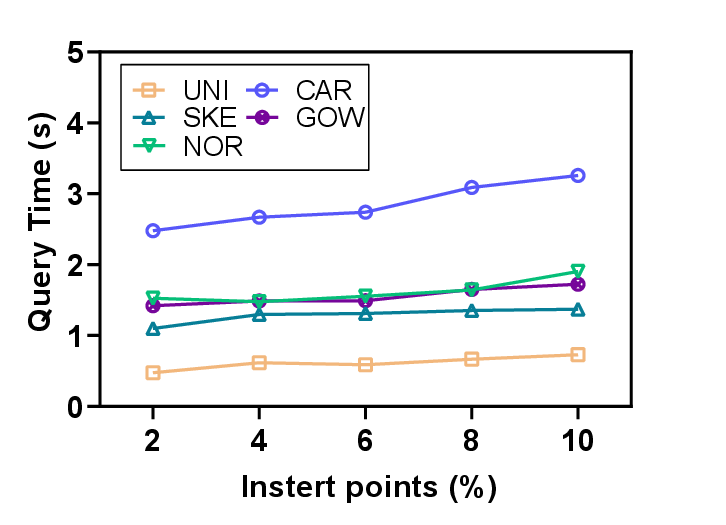}\label{fig:04_b}}
	\end{minipage}%
	\vspace{-3mm}
	\caption{Insertion and queries after insertions.}
	\label{fig04}
	\vspace{-5mm}
\end{figure}

{\bf Index update.} We further evaluated the update efficiency of \gls{idx}. Since deletion operations are relatively straightforward, their performance is not reported separately. When $n=20000$, we inserted 2\%, 4\%, 6\%, 8\%, and 10\% of additional data points, respectively. As shown in Fig.~\ref{fig04}(a), the update latency increases with the number of inserted points. 
Fig.~\ref{fig04}(b) illustrates that query latency increases across all datasets as more points are inserted.

%

\subsection{Secure Query Processing}
\begin{figure}[t]
	\centering
	\includegraphics[width=6.1cm]{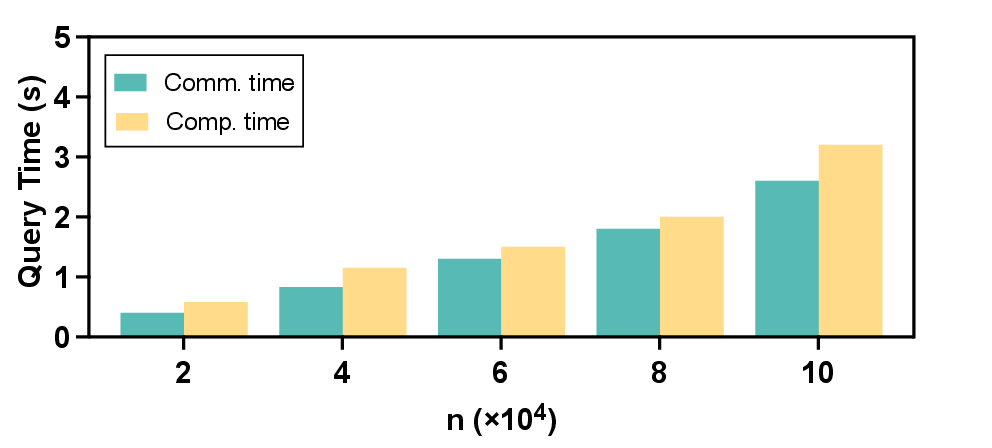}\vspace{-2mm}
	\caption{Communication Evaluation.} 
	\label{commu}
	\vspace{-5mm}
\end{figure}

{\bf\gls{RQ} Communication Evaluation.} To evaluate the communication overhead, we deployed the DSP and DAP on separate machines within the campus network. Fig. \ref{commu} illustrates communication time is slightly less than computation time, accounting for nearly half of the total runtime.

\begin{figure}[t]
	\begin{minipage}[h]{0.5\linewidth}
		\centering
		\subfigure[Query time]{\includegraphics[width=1.79in]{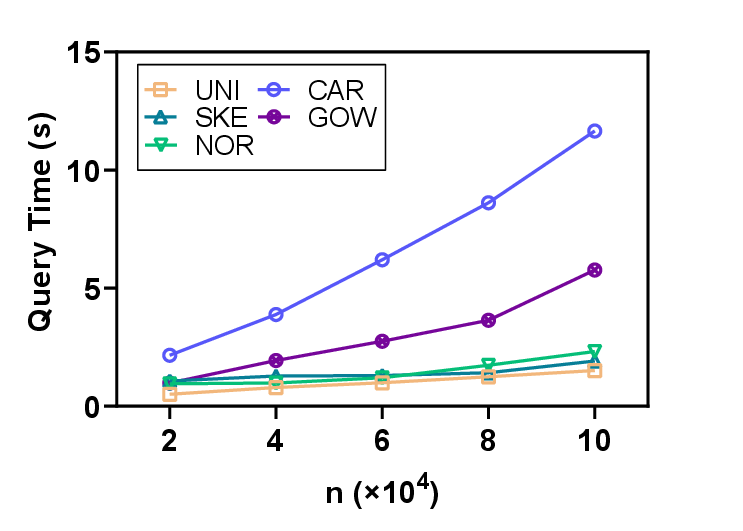}\label{fig:05_a}}\hspace{1mm}
	\end{minipage}%
	\begin{minipage}[h]{0.5\linewidth}
		\centering
		\subfigure[Recall]{\includegraphics[width=1.79in]{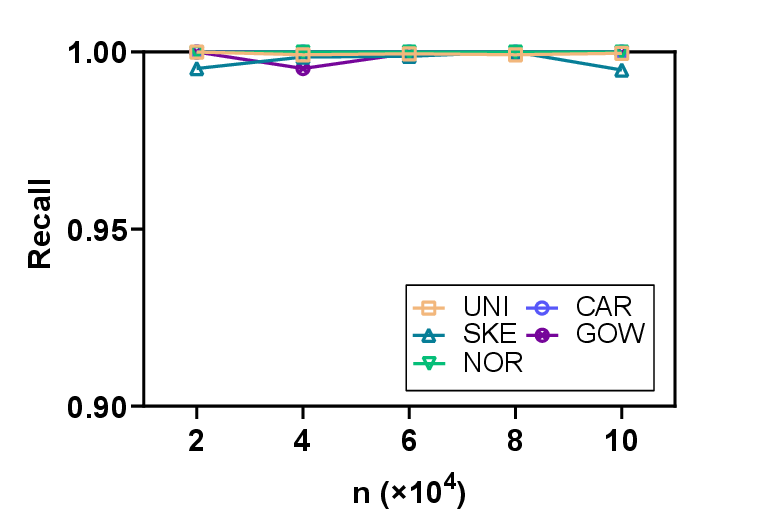}\label{fig:05_b}}
	\end{minipage}%
	\vspace{-3mm}
	\caption{Impact of $n$ on query effciency.}
	\label{fig05}
	\vspace{-5mm}
\end{figure}

{\bf Impact of different parameters.} We evaluated the impact of dataset size $n$ on query latency and recall. As shown in Fig.~\ref{fig05}(a), with increasing $n$, a fixed query size and aspect ratio correspond to a larger volume of covered data, leading to significant computational overhead. Overall, query latency increases with $n$ across all datasets. In Fig.~\ref{fig05}(b), we observe that recall remains relatively stable as $n$ grows, staying close to 1. This result validates the high accuracy of \gls{RQ} queries.

Table~\ref{tab08} presents the impact of varying $m$ on query performance across datasets. Each dataset shows a distinct optimal $m$, reflecting a trade-off between data distribution and homomorphic encryption overhead. Skewed datasets like GOW benefit from larger $m$ to reduce redundant bucket accesses in dense regions. Sparse datasets such as CAR also favor larger $m$ to balance partition count and block density. In contrast, uniformly distributed datasets perform better with smaller $m$, which promotes balanced partitioning and lowers predictor errors.

\begin{figure}[t]
	\begin{minipage}[h]{0.5\linewidth}
		\centering
		\subfigure[Impact of b]{\includegraphics[width=1.79in]{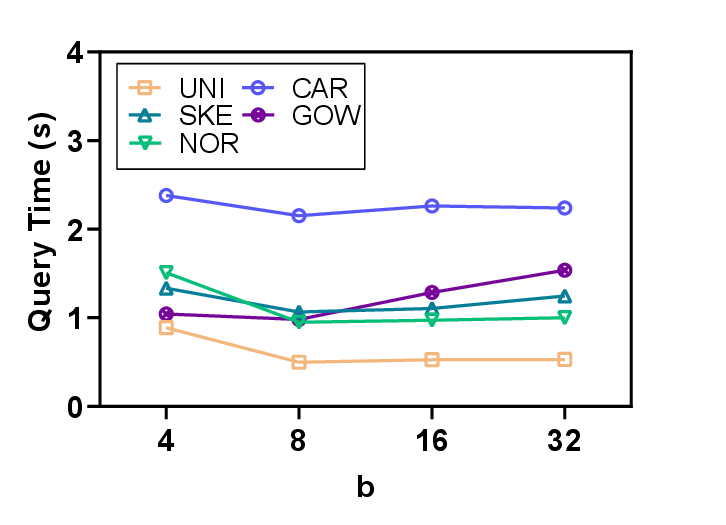}\label{fig:06_a}}
	\end{minipage}%
	\begin{minipage}[h]{0.5\linewidth}
		\centering
		\subfigure[Impact of K]{\includegraphics[width=1.79in]{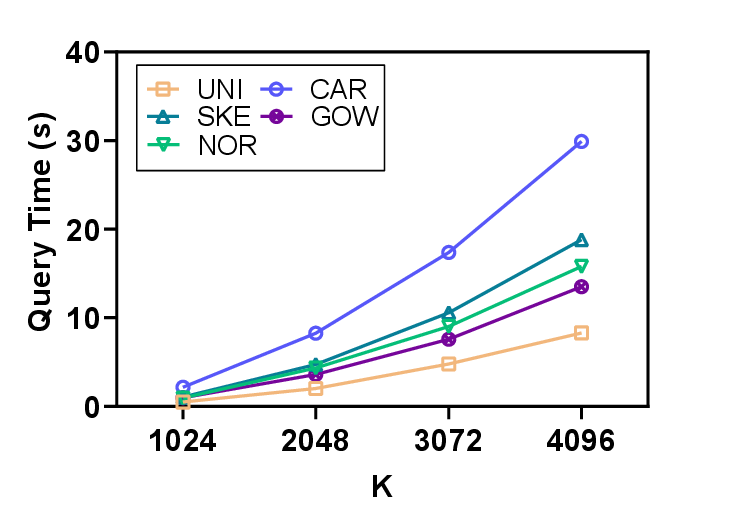}\label{fig:06_b}}
	\end{minipage}
	
	\begin{minipage}[h]{0.5\linewidth}
		\centering
		\subfigure[Impact of area]{\includegraphics[width=1.79in]{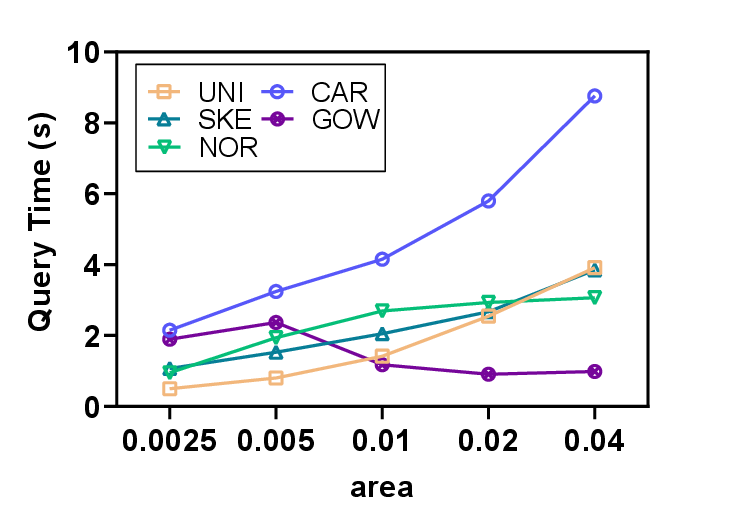}\label{fig:06_c}}
	\end{minipage}%
	\begin{minipage}[h]{0.5\linewidth}
		\centering
		\subfigure[Impact of radio]{\includegraphics[width=1.79in]{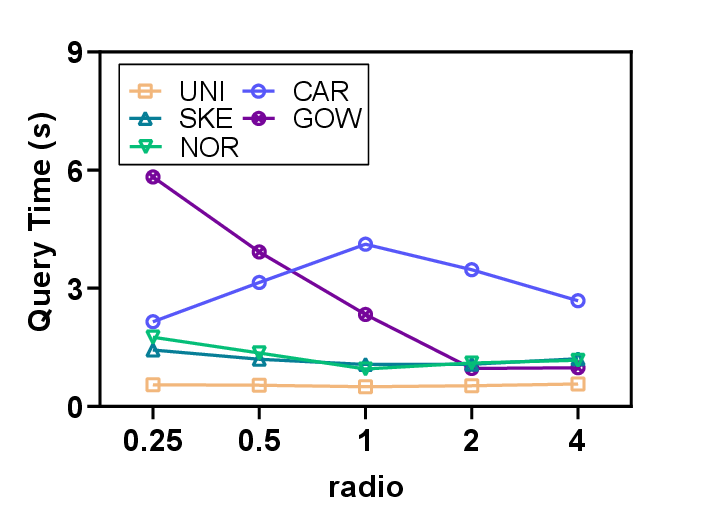}\label{fig:06_d}}
	\end{minipage}
	
	\vspace{-3mm}
	\caption{Query efficiency.}
	\label{fig06}
	\vspace{-5mm}
\end{figure}

As shown in Fig.~\ref{fig:06_a}, we analyze the impact of \(b\) on query efficiency. Across all datasets, query latency reaches its minimum at $b = 8$, yielding an average speedup of 37\% over \(b = 4\) and 21\% over \(b = 32\). This suggests that \(b = 8\) offers the best trade-off under Paillier, balancing bucket access frequency (too small \(b\) increases I/O) and scanning overhead (too large \(b\) amplifies homomorphic computations). Skewed datasets like GOW are more sensitive to \(b\), where suboptimal values either increase \gls{spe}-based retrievals or introduce redundant scans, both harming query efficiency.


Fig.~\ref{fig:06_b} presents the impact of Paillier key length $K$ on query efficiency. As $K$ increases, query latency exhibits an almost exponential growth trend. 
Furthermore, the bit expansion of ciphertexts leads to significant performance degradation in ciphertext protocols like SIC and SM, ultimately resulting in increased query latency.

Fig.~\ref{fig:06_c} shows the impact of query size on efficiency. For all datasets except GOW, query latency increases proportionally with the query range, as larger size cover more points and impose higher computational overhead on \gls{RQ}. GOW, due to its skewed distribution, exhibits a non-monotonic trend with an optimal query size; for example, latency at size 0.02 is 0.9 seconds, much lower than 2.3 seconds at size 0.005.

Fig.~\ref{fig:06_d} depicts the effect of aspect ratio on query latency under a fixed size of 0.0025. For synthetic datasets, latency decreases then increases with aspect ratio, minimizing at 1, reflecting uniform local distributions (especially Uniform dataset). In contrast, CAR shows the opposite trend with a minimum at 0.25, due to its skewed local distribution. GOW’s latency fluctuates irregularly, attributed to its larger query size (i.e., 0.04) and more severe skewness.

\begin{figure}[t]
	\begin{minipage}[h]{0.5\linewidth}
		\centering
		\subfigure[Query efficiency over UNI]{\includegraphics[width=1.83in]{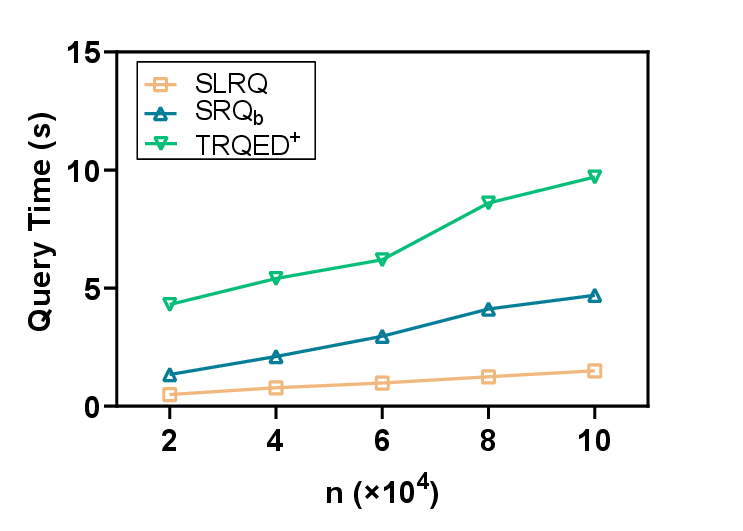}\label{fig:07_a}}\hspace{1mm}
	\end{minipage}%
	\begin{minipage}[h]{0.5\linewidth}
		\centering
		\subfigure[Query efficiency over GOW]{\includegraphics[width=1.79in]{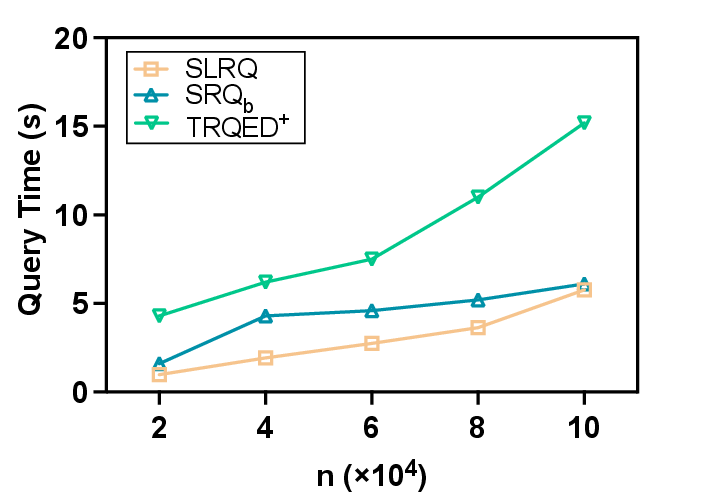}\label{fig:07_b}}
	\end{minipage}%
	\vspace{-3mm}
	\caption{Query efficiency (\gls{RQ} vs. TRQED$^{+}$ vs. SRQ$_{b}$).}
	\label{fig07}
	\vspace{-5mm}
\end{figure}

{\bf Comparing with prior schemes.}
Fig.~\ref{fig07} compares the query efficiency of \gls{RQ}, SRQ$_{b}$, and TRQED$^{+}$ on the synthetic UNI dataset and the real-world GOW dataset. 
Overall, \gls{RQ} consistently outperforms both TRQED$^{+}$ and SRQ$_{b}$. 
By leveraging learned indexes, \gls{RQ} rapidly predicts target regions, achieving higher efficiency than TRQED$^{+}$ and outperforming SRQ$_{b}$ on less extreme datasets, while providing stronger privacy guarantees.

\section{Related Work}
In recent years, privacy-preserving range queries over encrypted cloud data have attracted significant research interest. Some work relies on hardware-based solutions such as ORAM \cite{stefanov2018path}, but the practical efficiency of ORAM remains a major obstacle for real-world deployment. Consequently, alternative approaches focus on designing advanced data structures to reduce the computational overhead of privacy preservation. Among methods employing traditional data structures, Yang et al. \cite{yang2020achieving} proposed an encrypted R-tree with secure range query protocols, yet neither the baseline nor enhanced versions protect against access pattern leakage. Chen et al. \cite{chen2023efficient} combined Yao’s garbled circuits with R-trees for privacy-preserving range queries, but their approach incurs substantial computational cost due to extensive non-XOR gate operations per query. Cui et al. \cite{cui2024enabling} introduced multi-user privacy-preserving queries leveraging Prefix-aware encoding and SATree, but their scheme suffers from non-trivial encoding and hashing costs and lacks support for dynamic data updates. Zheng et al. \cite{zheng2024epset} built a pivot k-d tree based on Jaccard distances, implementing privacy-preserving queries via homomorphic encryption protocols. However, the k-d tree traversal leaks approximate query orientation, compromising access pattern privacy. Moreover, M$^{*}$-tree-based approaches \cite{cui2025towards} remain computationally expensive for privacy-preserving range queries.

Another line of research employs encoding techniques for privacy-preserving range queries. Liang et al. \cite{liang2024efficient} designed binary tree indexes by integrating range encoders with Additional Symmetric-Key Hidden Vector Encryption (ASHVE), yet this scheme inadequately protects both search and access pattern privacy. Wang et al. \cite{wang2025ppsksq} developed comparison protocols based on improved Symmetric Homomorphic Encryption (iSHE) combined with quadtree indexes to encode spatial range data. Despite employing random node ordering and dynamic tree depth to mitigate privacy risks, quadtree traversal still exposes access patterns. Similarly, Order-Revealing Encryption (ORE)-based indexes \cite{wang2025beyond} offer solutions for privacy-preserving range queries but face memory constraints within SGX enclaves, limiting scalability for large datasets.

Meanwhile, learned indexes have significantly enhanced database query performance~\cite{ferragina2020learned}. By integrating machine learning with traditional data structures, Peng et al.~\cite{yongxin2020study} proposed Learned KD-Trees, replacing classic KD-trees with neural networks to reduce search latency without compromising accuracy. Maharjan et al.~\cite{maharjan2023leanstore} incorporated learned indexes into \( B^{+} \)-trees, eliminating layer-wise traversal and substantially improving query speed. Similarly, Song et al.~\cite{song2023lifm} employed in-memory learned indexes within hierarchical \( B^{+} \)-trees to reduce I/O overhead and enable persistent indexing. Huang et al.~\cite{huang2024hlihp} introduced a hierarchical learned index enhanced by a precision correction model, while Qi et al.~\cite{qi2020effectively} developed a multi-layer perceptron-based index using rank-space sorting and recursive partitioning for efficient spatial indexing. Notably, existing work on learned indexes primarily targets plaintext data, with limited exploration in encrypted settings.

\section{Conclusion}
This paper presents \gls{RQ}, a secure and efficient range query scheme over encrypted data, built upon a novel learned index called \gls{idx}. \gls{idx} integrates homomorphic encryption with a hierarchical prediction architecture to enable data-aware query processing while preserving privacy. 
To protect search and access patterns, we design a series of secure sub-protocols leveraging permutation and obfuscation techniques, including noise padding and dummy bucket injection.
Our scheme ensures strong privacy guarantees across data, queries, results and access patterns. Extensive experiments on both real-world and synthetic datasets demonstrate that \gls{RQ} significantly outperforms existing solutions in query efficiency.

\section*{Acknowledgment}

This work was supported by the National Natural Science Foundation of China under Grants 62502185 and 62272123, and by the Wuxi Science and Technology Development Fund Project under Grant K20241028.

\bibliographystyle{IEEEtran}  

\bibliography{myref}

\end{document}